\documentclass[a4paper,11pt]{article}
\usepackage[utf8x]{inputenc}
\usepackage[left=3cm,right=3cm,top=3cm,bottom=3cm]{geometry}
\usepackage[linktoc=all,hidelinks]{hyperref}
\usepackage{cite}
\usepackage{booktabs}
\usepackage{amsmath}
\usepackage{amsfonts}
\usepackage{amssymb}
\usepackage{tensor}
\usepackage{slashed}
\usepackage{tikz}

\newcommand{\pd}{\partial}

\newcommand{\nn}{\nonumber}

\newcommand{\cD}{\mathcal{D}}

\newcommand{\cM}{\mathcal{M}}
\newcommand{\cN}{\mathcal{N}}

\newcommand{\cV}{\mathcal{V}}

\newcommand{\SM}{S^\text{M}}
\newcommand{\SNM}{S^\text{NM}}
\newcommand{\tSNM}{\widetilde{S}^\text{NM}}

\DeclareMathOperator{\gh}{gh}

\newcommand{\cc}{\text{c.c.}}

\renewcommand{\tilde}{\widetilde}

\newcommand{\id}{\mathbf{1}}
\DeclareMathOperator{\ch}{ch}
\DeclareMathOperator{\tr}{tr}
\DeclareMathOperator{\Tr}{Tr}

\newcommand{\AM}{\widehat{A}(\cM)}

\begin{document}
\numberwithin{equation}{section}

\thispagestyle{empty}
\begin{center}

\vspace*{50pt}
{\LARGE \bf
On the quantisation and anomalies of\\\vspace{0.3cm}
antisymmetric tensor-spinors}

\vspace{30pt}
{Victor Lekeu${}^{\,a}$ and Yi Zhang${}^{\, a,b}$}

\vspace{10pt}
\texttt{victor.lekeu@aei.mpg.de, yi.zhang@aei.mpg.de}

\vspace{20pt}
\begin{enumerate}
\item[${}^a$] {\sl \small
Max-Planck-Institut für Gravitationsphysik (Albert-Einstein-Institut)\\
Am Mühlenberg 1, 14476 Potsdam, Germany}
\item[${}^b$] {\sl \small
Institut de Physique Théorique, Université Paris Saclay, CNRS, CEA\\
91191 Gif-sur-Yvette Cedex, France}
\end{enumerate}

\vspace{50pt}
{\bf Abstract} 
\end{center}

\noindent

We perform the quantisation of antisymmetric tensor-spinors (fermionic $p$-forms) $\psi^\alpha_{\mu_1 \dots \mu_p}$ using the Batalin-Vilkovisky field-antifield formalism. Just as for the gravitino ($p=1$), an extra propagating Nielsen-Kallosh ghost appears in quadratic gauges containing a differential operator. The appearance of this `third ghost' is described within the BV formalism for arbitrary reducible gauge theories. We then use the resulting spectrum of ghosts and the Atiyah-Singer index theorem to compute gravitational anomalies.

\newpage

\setcounter{tocdepth}{2}
\tableofcontents


\section{Introduction}

We consider in this paper the quantisation of antisymmetric tensor-spinors. These are fermionic fields of the form $\psi^\alpha_{\mu_1\mu_2 \dots \mu_p}$, where $\alpha$ is a spinor index and the $\mu_i$ are spacetime indices, which are totally antisymmetric in their spacetime indices:
\begin{equation}
\psi^\alpha_{\mu_1\mu_2 \dots \mu_p} = \psi^\alpha_{[\mu_1\mu_2 \dots \mu_p]}\, .
\end{equation}
We will also call them `fermionic $p$-forms'. The free action for such a field in flat spacetime is a direct generalisation of the Rarita-Schwinger action for a fermionic one-form $\psi^\alpha_\mu$ and reads \cite{Buchbinder:2009pa,Zinoviev:2009wh,Campoleoni:2009gs}
\begin{equation}\label{eq:actionintro}
    S_0[\psi] = - (-1)^{\frac{p(p-1)}{2}} \, \int \! d^D\!x\; \bar{\psi}_{\mu_1\mu_2\dots \mu_p} \,\gamma^{\mu_1\mu_2\dots \mu_p \nu \rho_1\rho_2\dots \rho_p} \,\pd_\nu \psi_{\rho_1\rho_2\dots \rho_p}\, .
\end{equation}
This action is invariant under some \emph{reducible} gauge symmetries, i.e.~with `gauge-for-gauge' transformations. They are
\begin{equation}
\delta \psi = d \Lambda^{(p-1)}\, , \quad \delta \Lambda^{(p-1)} = d\Lambda^{(p-2)}\, , \quad \dots \, , \quad \delta \Lambda^{(1)} = d\Lambda^{(0)}
\end{equation}
in differential form notation (with a spectator spinor index). Here, each parameter $\Lambda^{(k)}$ is an antisymmetric tensor-spinor of rank $k$. This reducibility introduces well-known subtleties upon quantisation, which we will tackle using the powerful Batalin-Vilkovisky (BV) field-antifield formalism \cite{Batalin:1981jr,Batalin:1984jr}.

Our motivation for examining these fields is twofold. First, fermionic two-forms appear in the exotic $\cN = (4,0)$ and $\cN = (3,1)$ maximally supersymmetric multiplets in six dimensions \cite{Strathdee:1986jr}, which have been conjectured by Hull to play a role in some strongly coupled regimes of maximal supergravity \cite{Hull:2000zn,Hull:2000rr} and have attracted renewed interest in recent years \cite{Borsten:2017jpt,Henneaux:2017xsb,Henneaux:2018rub,Minasian:2020vxn,Bertrand:2020nob,Gunaydin:2020mod,Galati:2020vqp,Cederwall:2020dui}. In particular, gravitational anomalies for these exotic multiplets were computed in \cite{Minasian:2020vxn}, but some assumptions were required since the precise ghost structure for the fermionic two-form was unknown at the time. One of the goals of this paper is to fill that gap. Another, more remote motivation for looking at these types of fields comes from considerations of dual gravity \cite{Hull:2000zn,Hull:2000rr,West:2001as,Hull:2001iu}, where (in the linearised regime) the graviton is dualised to a $[D-3,1]$-type mixed-symmetry tensor. A supersymmetric, manifestly covariant model in which this field finds a partner is still lacking, however, and a fermionic $p$-form field would be the natural candidate (see \cite{Townsend:1979yv} for an early attempt at dualising fermionic fields, and \cite{Bunster:2012jp,Lekeu:2018kul} for related considerations in the prepotential formalism).

The structure and main results of this paper are as follows.
\begin{itemize}
\item Section \ref{sec:reviewNK} starts with a short review of the quantisation procedure of irreducible gauge theories in the BV formalism \cite{Batalin:1981jr}, i.e., when there are no `gauge-for-gauge' transformations. We put a special emphasis on quadratic gauges containing a differential operator: there, a third propagating ghost appears, as was first described within this formalism in a manifestly local way by Batalin and Kallosh in \cite{Batalin:1983ar}. Then, this procedure is applied to the quantisation of the free Rarita-Schwinger field, where this propagating third ghost is known as the Nielsen-Kallosh ghost \cite{Nielsen:1978mp,Kallosh:1978de}.
\item In section \ref{sec:BVred}, we discuss in the BV formalism the appearance of the `third ghost' for quadratic gauges in an arbitrary gauge theory. This generalises a result of \cite{Batalin:1983ar} to the reducible case and is one of the main results of this paper. It should be emphasised that this result is valid beyond the simple action and gauge symmetries for the fermionic $p$-form described above: we allow for non-abelian gauge algebras, on-shell closure, etc. These subtleties are all packaged in the explicit form of the `minimal BV action' for the model at hand, which always exists and which we keep arbitrary.
\item Section \ref{sec:pforms} turns to the quantisation of free fermionic $p$-form fields, using the general results of the previous section. (In the words of \cite{Batalin:1984jr} however, this is `like cracking nuts with a sledgehammer'.) This is done both in the usual delta-function gauge-fixing and in the Gaussian gauge-fixing where a generalised Nielsen-Kallosh ghost appears; propagators and BRST transformations are also discussed in both schemes. Explicit details are given only in the two-form case, but the generalisation to higher form degree poses no difficulty. We maintain manifest locality and covariance throughout.
\item Finally, in section \ref{sec:anomalies} we compute the gravitational anomaly of a chiral fermionic $p$-form in dimensions $D = 4m + 2$. This is done using the ghost spectrum found in the previous section and applying the Atiyah-Singer index theorem \cite{Atiyah:1970ws,Atiyah:1971rm}, following the methods developed in the classic papers \cite{Alvarez-Gaume:1983ihn,Alvarez-Gaume:1983ict,Alvarez-Gaume:1984zlq,Alvarez:1984yi}. We describe the general procedure in detail and display the results in dimensions $D = 2$, $6$ and $10$ in tables \ref{tab:D2}, \ref{tab:D6} and \ref{tab:D10}. An intriguing result is that in dimensions $D \geq 6$, the anomaly of a chiral fermionic $p$-form matches that of a $(D-p-1)$-form; it would be very interesting to use this fact to attempt to build new anomaly-free models.
\end{itemize}
We should mention an important caveat related to the computation of the gravitational anomaly: to the best of our knowledge, there is currently no model that couples consistently a fermionic $p$-form to dynamical gravity. It can be hoped that this difficulty will be resolved in the future, perhaps by including (an infinite number of) other fields\footnote{Something even more exotic should happen in the $D=6$, $\cN = (4,0)$ or $(3,1)$ theories, if they exist, since they contain no metric at all. There, one should probably take the vanishing of the gravitational anomaly as a criterion selecting on which background manifolds these theories can be formulated consistently in certain regimes.}. However, since the anomaly computations of section \ref{sec:anomalies} are solely based on the ghost spectrum and not on the specific form of the action, we are confident that these results will survive such future developments.

\section{Review: the Nielsen-Kallosh ghost}\label{sec:reviewNK}

We begin with a short review of BV quantisation of irreducible gauge theories and apply it to the free gravitino field. This will be generalised to arbitrary reducible gauge theories in section \ref{sec:BVred}, and those results will be applied to antisymmetric tensor-spinors in section \ref{sec:pforms}. 

\subsection{In the Batalin-Vilkovisky formalism}\label{sec:BVirr}

We start from an action $S_0[\varphi^i]$ depending on fields $\varphi^i$, invariant under some gauge invariances $\delta \varphi^i = R^i_\alpha \Lambda^\alpha$ with generator $R^i_\alpha$ and parameter $\Lambda^\alpha$. Gauge invariance is equivalent to the Noether identities
\begin{equation}
    \frac{\delta^R S_0}{\delta \varphi^i} R^i_\alpha = 0\, .
\end{equation}
We follow the notation and conventions of \cite{Henneaux:1992ig,Gomis:1994he}; in particular, a contracted index includes space-time integration and the superscript $R$ (resp.~$L$) indicates that the derivative is acting from the right (resp. left).

We assume in this section that the theory is irreducible, so there are no `gauge-for-gauge' transformations: the gauge generators $R^i_\alpha$ are independent on-shell. We will also assume the usual regularity conditions on $S_0$ throughout this paper; these are detailed in the reviews \cite{Henneaux:1992ig,Gomis:1994he}. The fields and gauge parameters are allowed to be bosonic or fermionic: their Grassmann parity is written as $\epsilon(\varphi_i) \equiv \epsilon_i$ and $\epsilon(\Lambda^\alpha) \equiv \epsilon_\alpha$. The gauge-fixing condition will be written $\chi^\alpha(\varphi) = 0$; the function $\chi^\alpha(\varphi)$ has the same index structure and Grassmann parity as the gauge parameter $\Lambda^\alpha$.

In the field-antifield formalism of Batalin and Vilkovisky \cite{Batalin:1981jr,Batalin:1984jr}, the space of fields is extended to include a ghost field $c^\alpha$ corresponding to the gauge parameter $\Lambda^\alpha$, and the antifields $\varphi^*_i$ and $c^*_\alpha$. Ghost number assignments and Grassmann parities are collected in table \ref{tab:irreducible}. We will denote the set of fields collectively by $\Phi^I$, and the antifields by $\Phi^*_I$.
\begin{table}
    \centering
    \begin{tabular}{c| c c c c | c c c}
         & $\varphi^i$ & $c^\alpha$ & $c'^\alpha$ & $b^\alpha$ & $S_0$, $\SM$, $\SNM$ & $\Psi$ & $\chi^\alpha$ \\ \midrule
        $\gh$ & $0$ & $1$ & $-1$ & $0$ & $0$ & $-1$ & $0$ \\[\defaultaddspace]
        $\epsilon$ & $\epsilon_i$ & $\epsilon_\alpha+1$ & $\epsilon_\alpha+1$ & $\epsilon_\alpha$ & $0$ & $1$ & $\epsilon_\alpha$
    \end{tabular}
    \caption{Ghost numbers and Grassmann parities of the various objects appearing in the irreducible case. Antifields have ghost number given by $\gh(\Phi^*_I) = - \gh(\Phi^I) - 1$ and the opposite parity, $\epsilon(\Phi^*_I) = \epsilon(\Phi^I) + 1$.}
    \label{tab:irreducible}
\end{table}
The action $S_0[\varphi^i]$ is then extended to the minimal BV action $\SM[\varphi^i, c^\alpha; \varphi^*_i, c^*_\alpha]$
depending on the original fields $\varphi^i$ but also on the ghost field $c^\alpha$ and their antifields $\varphi^*_i$, $c^*_\alpha$. It is a ghost number zero, even functional that should be a proper solution of the classical master equation
\begin{equation}
(\SM, \SM) = 0\, ,
\end{equation}
where the antibracket $(\, \cdot \, , \, \cdot \,)$ is defined as
\begin{equation}
    (X,Y) = \frac{\delta^R X}{\delta \Phi^I}\frac{\delta^L Y}{\delta \Phi^*_I} - \frac{\delta^R X}{\delta \Phi^*_I}\frac{\delta^L Y}{\delta \Phi^I}\, .
\end{equation}
Moreover, it should reduce to the original action when the antifields are set to zero:
\begin{equation}
S_0[\varphi^i] = \SM[\Phi, \Phi^* = 0]\, .
\end{equation}
These two conditions completely determine $\SM$, which always exists; it starts with
\begin{equation}
S^\text{M}[\varphi^i, c^\alpha; \varphi^*_i, c^*_\alpha] = S_0[\varphi] + \varphi^*_i R^i_\alpha c^\alpha + \dots
\end{equation}
and the omitted terms carry the explicit information about the gauge algebra, on-shell closure, etc. We refer to \cite{Henneaux:1992ig,Gomis:1994he} for pedagogical reviews.

To gauge-fix the theory, one further extends the space of fields by adding a trivial pair of fields $(c'^\alpha, b^\alpha)$ of ghost numbers $-1$ and $0$ respectively, along with their antifields\footnote{In some cases, it can be more convenient to take these fields with down indices instead, and we will sometimes do this in the following. Note also that $c'$ is often written $\bar{c}$; however, since we will be dealing with fermionic theories in the applications, this could be confused with the Dirac conjugate and we will stick with the prime notation in this paper.}. The minimal action $\SM$ is then extended to the non-minimal
\begin{equation}\label{eq:SNMirr}
    \SNM = \SM[\varphi^i, c^\alpha; \varphi^*_i, c^*_\alpha] + c'^*_\alpha b^\alpha\, ,
\end{equation}
which still satisfies the master equation. The antifields are then eliminated according to the formula
\begin{equation}\label{eq:gffermion}
\Phi^*_I = \frac{\delta \Psi}{\delta \Phi^I}\, ,
\end{equation}
where $\Psi(\Phi)$ is an odd functional of ghost number $-1$ depending on the fields only, called the gauge-fixing fermion. It does not matter whether one uses left or right derivatives in \eqref{eq:gffermion}. If $\Psi$ is well-chosen, the resulting action is properly gauge-fixed and possesses well-defined propagators.

The simplest example is delta-function gauge-fixing: here, one can simply take the gauge-fixing fermion as
\begin{equation}
    \Psi_\delta = c'_\alpha \chi^\alpha(\varphi)\, .
\end{equation}
This gives the gauge-fixed action
\begin{align}
S_\delta[\varphi^i, c^\alpha, c'_\alpha, b_\alpha] &\equiv \SNM\left[ \Phi^I, \Phi^*_I = \frac{\delta \Psi_\delta}{\delta \Phi^I} \right] \\
&= \SM\left[ \varphi^i, c^\alpha; \varphi^*_i = c'_\alpha \frac{\delta^R \chi^\alpha}{\delta \varphi^i} , c^*_\alpha = 0 \right] + \chi^\alpha(\varphi) b_\alpha\, . \label{eq:deltairr}
\end{align}
The field $b^\alpha$ is auxiliary and enforces the gauge-fixing constraint $\chi^\alpha(\varphi) = 0$. The fields $c^\alpha$ and $c'_\alpha$ are the usual Faddeev-Popov ghosts. Note however that formula \eqref{eq:deltairr} is also correct for e.g.~theories with open algebras where the usual Faddeev-Popov procedure cannot be applied; these subtleties only appear in the explicit form of $\SM$.

It can also be convenient to use Gaussian gauge-fixing, with a gauge-breaking term in the action of the form $\chi_\alpha(\varphi) M^{\alpha\beta} \chi_\beta(\varphi)$ with some non-degenerate matrix $M$. This can be obtained by including terms linear in the auxiliary fields in $\Psi$:
\begin{equation}\label{eq:usualgaussianpsi}
    \Psi = c'_\alpha \chi^\alpha(\varphi) + \frac{1}{2} c'_\alpha (M^{-1})^{\alpha\beta} b_\beta\, ,
\end{equation}
giving
\begin{align}\label{eq:usualgaussianS}
S[\varphi^i, c^\alpha, c'_\alpha, b_\alpha] = \SM\left[ \varphi^i, c^\alpha; \varphi^*_i = c'_\alpha \frac{\delta^R \chi^\alpha}{\delta \varphi^i} , c^*_\alpha = 0 \right] + \chi^\alpha(\varphi) b_\alpha + \frac{1}{2} (M^{-1})^{\alpha\beta} b_\beta b_\alpha \, .
\end{align}
Here, $b^\alpha$ is a simple auxiliary field appearing quadratically in the action. Eliminating it using its own equation of motion yields the looked-after gauge-breaking term $\chi_\alpha M^{\alpha\beta} \chi_\beta$. However, in some applications one would like this term to contain a differential operator, $M = \cD$. Then, the above procedure is problematic since the non-local object $\cD^{-1}$ appears in the gauge-fixing fermion \eqref{eq:usualgaussianpsi} and the action \eqref{eq:usualgaussianS}.

It is well-known that gauge-breaking terms of this form lead to a third propagating ghost, the Nielsen-Kallosh ghost \cite{Nielsen:1978mp,Kallosh:1978de} (the first two ghosts being the usual Faddeev-Popov ghosts $c^\alpha$ and $c'^\alpha$). This was first described within this formalism, while maintaining manifest locality throughout, by Batalin and Kallosh in reference \cite{Batalin:1983ar} as we review shortly now. The third ghost is nothing but the $b^\alpha$ field, which stops being auxiliary and propagates with kinetic operator $\cD$.

The trick is to use the freedom to do a canonical transformation, which preserves the antibracket and maps solutions of the master equation to solutions, and only after that replace the antifields using a gauge-fixing fermion. In the simple case where the gauge condition $\chi^\alpha$ only depends on the original fields $\varphi^i$, the canonical transformation reads
\begin{align}
    b^\alpha &\rightarrow \tilde{b}^\alpha = b^\alpha - \chi^\alpha(\varphi) \nn\\
    \varphi^*_i &\rightarrow \tilde{\varphi}^*_i = \varphi^*_i + b^*_\alpha \frac{\delta^R \chi^\alpha}{\delta \varphi^i}\, .\label{eq:canonicalirr}
\end{align}
(the gauge condition $\chi^\alpha$ is also allowed to depend on the ghost fields $c^\alpha$, $c'^\alpha$ or $b^\alpha$, in which case the canonical transformation is more complicated; see \cite{Batalin:1983ar}), with other variables unchanged.\footnote{To check that this transformation is canonical, compute
\begin{align}
    \tilde{\varphi}^*_i d \tilde{\varphi}^i + \tilde{b}^*_\alpha d \tilde{b}^{\alpha} &=  \left( \varphi^*_i + b^*_\alpha \frac{\delta^R \chi^\alpha}{\delta \varphi^i} \right) d\varphi^i + b^*_\alpha d ( b^\alpha - \chi^\alpha(\varphi) ) = \varphi^*_i d \varphi^i + b^*_\alpha d b^{\alpha} \nn\, ,
\end{align}
which is the field-antifield analogue of the condition $p'_i dq'^i = p_i dq^i$ in classical mechanics. Another way is to notice that this transformation is generated by $F = b^*_\alpha \chi^\alpha(\varphi)$ via the antibracket, i.e.~takes the form
\begin{equation}
    \Phi^I \rightarrow \Phi^I + (F, \Phi^I) \, , \quad \Phi^*_I \rightarrow \Phi^*_I + (F, \Phi^*_I) \, , \quad F=b^*_\alpha \chi^\alpha(\varphi)\, .\nn
\end{equation}}
This maps the non-minimal action \eqref{eq:SNMirr} to
\begin{align}
    \tSNM = \SM\left[\varphi^i, c^\alpha; \varphi^*_i + b^*_\alpha \frac{\delta^R \chi^\alpha}{\delta \varphi^i} , c^*_\alpha\right] + c'^*_\alpha ( b^\alpha - \chi^\alpha(\varphi) )\, ,
\end{align}
which still satisfies the master equation since the transformation is canonical. Using the gauge-fixing fermion
\begin{equation}
    \Psi_\text{G} = \frac{1}{2} c'^\alpha \cD_{\alpha\beta}(\varphi) \left( \chi^\beta(\varphi) + b^\beta \right)\, .
\end{equation}
now gives
\begin{align}
    S_\text{G} &\equiv \tSNM\left[ \Phi^I, \Phi^*_I = \frac{\delta \Psi_\text{G}}{\delta \Phi^I} \right] \\
    &= \SM \left[\varphi^i, c^\alpha ; c'^\alpha \cD_{\alpha\beta} \frac{\delta^R \chi^\beta}{\delta \varphi^i} + \frac{1}{2} c'^\alpha \frac{\delta^R \cD_{\alpha\beta}}{\delta \varphi^i}\left( \chi^\beta + b^\beta \right) (-1)^{\epsilon_i \epsilon_\beta} , 0 \right] \nonumber \\
    &\quad - \frac{1}{2} \cD_{\alpha\beta} \,\chi^\beta \chi^\alpha + \frac{1}{2} \cD_{\alpha\beta} b^\beta b^\alpha\, .
\end{align}
This action contains the desired gauge-breaking term $\cD_{\alpha\beta} \,\chi^\beta \chi^\alpha$, along with a quadratic term in $b^\alpha$. This construction is most relevant when the operator $\cD_{\alpha\beta}$ is field dependent: then, $b^\alpha$ is coupled to the other fields (including the ghosts $c^\alpha$ and $c'^\alpha$) and cannot be ignored in Feynman diagram computations.

\subsection{Quantisation of the Rarita-Schwinger Lagrangian}\label{sec:RSquant}

As an example, we apply in this section the field-antifield method described above to the quantisation of the free spin $3/2$ field $\psi_\mu^\alpha$ ($\mu$ is a space-time index and $\alpha$ a spinor index). We will use Dirac spinors to avoid dimension-dependent discussions of chirality and/or reality conditions, but these can be included without difficulty. We are in flat Minkowski spacetime here and in section \ref{sec:pforms}. Our spinor conventions are as in the textbook \cite{Freedman:2012zz}.

The action and gauge invariances are
\begin{equation}
    S_0[\psi] = -\int \! d^D\!x\, \bar{\psi}_\mu \gamma^{\mu\nu\rho} \pd_\nu \psi_\rho\, , \quad \delta \psi_\mu^\alpha = \pd_\mu \Lambda^\alpha\, ,
\end{equation}
where the bar denotes the usual Dirac conjugate, $\bar{\psi}_\mu \equiv i (\psi_\mu)^\dagger \gamma^0$. We will impose the gauge condition
\begin{equation} \label{eq:gaugecondRS}
    \chi(\psi) \equiv \gamma^\mu \psi_\mu = 0\, .
\end{equation}

In the minimal sector, there is the field $\psi_\mu^\alpha$, the ghost $c^\alpha$ corresponding to the gauge parameter $\Lambda^\alpha$, and their antifields $\psi^{*\mu}_\alpha$, $c^*_\alpha$. Notice that the antifields carry naturally an index down, so they transform as conjugate spinors under Lorentz transformations. The minimal BV action is
\begin{align}
    \SM &= \int \! d^D\!x\, \left( - \frac{1}{2} \bar{\psi}_\mu \gamma^{\mu\nu\rho} \pd_\nu \psi_\rho + \psi^{*\mu}_\alpha \pd_\mu c^\alpha + \cc \right) \\
    &= \int \! d^D\!x\, \left( -\bar{\psi}_\mu \gamma^{\mu\nu\rho} \pd_\nu \psi_\rho + \psi^{*\mu} \pd_\mu c + (\pd_\mu \bar{c}) \bar{\psi}^{*\mu} \right)\, ,
\end{align}
where in the second line we suppressed the spinor indices and introduced the notation 
\begin{equation}\label{eq:barconjugate}
\bar{\psi}^{*\mu} \equiv i \gamma^0 (\psi^{*\mu})^\dagger    
\end{equation}
for the `Dirac conjugate' of a conjugate (index-down) spinor. With this notation, we have $\bar{\bar{\chi}} = +\chi$ for any spinor or conjugate spinor $\chi$, and the property $(ab)^\dagger = + \bar{b}\bar{a}$ for any conjugate spinor $a$ and spinor $b$.

For the non-minimal sector, one adds one pair of spinors $( c'^\alpha, b^\alpha )$ and their antifields. The ghost numbers are given in table \ref{tab:RSghostnumbers}.
\begin{table}
\centering
\begin{tabular}{c|cccc|cccc}
& $\psi^\alpha_\mu$ & $c^\alpha$ & $c'^\alpha$ & $b^\alpha$ & $\psi^{*\mu}_\alpha$ & $c^*_\alpha$ & $c'^*_\alpha$ & $b^*_\alpha$ \\ \midrule
$\gh$ & 0 & 1 & -1 & 0 & -1 & -2 & 0 & -1
\end{tabular}
\caption{The ghost numbers of the fields and antifields appearing in the quantisation of the Rarita-Schwinger Lagrangian.}
\label{tab:RSghostnumbers}
\end{table}
Grassmann parity is ghost number \emph{plus one} modulo two, since we have a fermionic theory and take the convention where degrees add up when determining signs. In particular, $c$ and $c'$ are bosonic (commuting) spinors, while $b$ has the correct spin-statistics.
The non-minimal action, adding the trivial pair, is simply
\begin{align}
    \SNM &= \SM + \int \! d^D\!x\, \left(  c'^*_\alpha b^\alpha +\cc \right) \\
    &= \int \! d^D\!x\, \left( -\bar{\psi}_\mu \gamma^{\mu\nu\rho} \pd_\nu \psi_\rho + \psi^{*\mu} \pd_\mu c + (\pd_\mu \bar{c}) \bar{\psi}^{*\mu} + c'^* b + \bar{b}\bar{c}'^* \right)\, .
\end{align}

\paragraph{Delta-function gauge-fixing.}

For $\delta$-function gauge-fixing, one takes the gauge-fixing fermion
\begin{equation}
    \Psi_{\delta}= \int \! d^D\!x\, \left( \bar{c}'\chi(\psi) + \cc \right) = \int \! d^D\!x\, \left( \bar{c}'\gamma^\mu \psi_\mu - \bar{\psi}_\mu \gamma^\mu c' \right)\,,
\end{equation}
which gives the gauge-fixed action
\begin{align}
    S_\delta[\Phi^I] &= \SNM\left[\Phi^I, \Phi^*_I = \frac{\delta \Psi_\delta}{\delta \Phi^I}\right] \\
    &= \int \! d^D\!x\, \left( -\frac{1}{2} \bar{\psi}_\mu \gamma^{\mu\nu\rho} \pd_\nu \psi_\rho + \bar{c}'\, \gamma^\mu \pd_\mu c + \bar{b} \gamma^\mu \psi_\mu + \cc \right)\, \\
    &= \int \! d^D\!x\, \left( -\bar{\psi}_\mu \gamma^{\mu\nu\rho} \pd_\nu \psi_\rho + \bar{c}'\, \gamma^\mu \pd_\mu c + \bar{c}\, \gamma^\mu \pd_\mu c' + \bar{b} \gamma^\mu \psi_\mu - \bar{\psi}_\mu \gamma^\mu b \right)\, .\label{eq:RSgaugefixeddelta}
\end{align}
The auxiliary field $b$ enforces the gauge-fixing condition $\gamma^\mu \psi_\mu = 0$. Using this condition, the kinetic term for $\psi_\mu$ reduces to $-\bar{\psi}^\mu \slashed{\pd} \psi_\mu$.

\paragraph{Gaussian gauge-fixing.}

We now want to produce the Gaussian gauge-breaking term 
\begin{equation}
    \xi \,\bar{\chi}(\psi) \,\slashed{\pd}\, \chi(\psi) = - \xi\, \bar{\psi}_\mu \gamma^\mu \gamma^\nu \gamma^\rho \pd_\nu \psi_\rho
\end{equation}
with an arbitrary parameter $\xi\neq 0$. As indicated above, this is done by the canonical transformation
\begin{align}
    b \rightarrow b - \chi(\psi) \, , \quad \psi^{*\mu} \rightarrow \psi^{*\mu} + b^* \frac{\delta \chi}{\delta \psi_\mu} = \psi^*_\mu + b^* \gamma_\mu
\end{align}
(and similarly for the Dirac conjugates), which gives the
non-minimal action
\begin{align}
    \tSNM &= \int \! d^D\!x\, \left( - \frac{1}{2} \bar{\psi}_\mu \gamma^{\mu\nu\rho} \pd_\nu \psi_\rho + (\psi^{*\mu} + b^* \gamma^\mu) \pd_\mu c + c'^* (b - \chi) + \cc \right) \, .
\end{align}
Eliminating the antifields by means of the gauge-fixing fermion
\begin{equation}
    \Psi_\text{G} = - \frac{\xi}{2} \int \! d^D\!x\; \bar{c}' \, \slashed{\pd} \left( \chi(\psi) +  b  \right) + \cc
\end{equation}
then produces
\begin{equation}\label{eq:RSgaussian}
S_\text{G} = \int \! d^D\!x\, \left( - \bar{\psi}_\mu \gamma^{\mu\nu\rho} \pd_\nu \psi_\rho + \xi \, \bar{\chi} \slashed{\pd} \chi - \xi\left( \bar{c}'\, \Box c + \bar{c}\, \Box c' \right) - \xi\,  \bar{b} \slashed{\pd} b  \right)\, .
\end{equation}
The field $b$ is now a propagating spin $1/2$ field. Note that the ghosts $c$, $c'$ count for four, since they come with the second-order $\Box = \slashed{\pd}\slashed{\pd}$ as kinetic operator.\footnote{This can be `undoubled' with a well-known trick (cf.~for example the textbook \cite{Siegel:1999ew}, exercise VIA4.2). Introduce a Lagrange multiplier $\lambda$, of ghost number $-1$, to impose the equation $\slashed{\pd} c = f$ with $f$ a new spinor of ghost number 1. This gives the equivalence $\bar{c}'\, \Box c  \sim \bar{c}' \slashed{\pd} f + \bar{\lambda} ( \slashed{\pd}c - f )$. This is diagonalised by the non-local triangular change of variables $c \rightarrow c + \slashed{\pd}^{-1} f$, with final result $\bar{c}'\, \Box c  \sim \bar{c}' \slashed{\pd} f + \bar{\lambda} \slashed{\pd}c$ featuring four fields with a first-order Lagrangian instead of two with a second-order one.} The field $b$, being of ghost number zero, has the correct spin-statistics. This gives indeed the requisite number of ghosts: $3 = 4 -1$.

In the action \eqref{eq:RSgaussian}, the Nielsen-Kallosh ghost $b$ is decoupled. However, in supergravity the operator $\slashed{\pd}$ appearing in the gauge-breaking term is covariantised and contains the vielbein and the spin-connection. Then, the field $b$ couples to the other fields and ghosts of the theory \cite{Nielsen:1978mp,Kallosh:1978de,Batalin:1983ar}.

\section{The third ghost in reducible theories}\label{sec:BVred}

In this section, we present a simple generalisation of the procedure of \cite{Batalin:1983ar} reviewed in section \ref{sec:BVirr} to the reducible case, i.e.~in the presence of `gauge-for-gauge' transformations. This gives a mechanism within the BV formalism detailing the appearance of a third ghost in quadratic gauges for an arbitrary reducible gauge theory.

\subsection{First-stage reducible}\label{sec:BVredfirststage}

We consider an action $S_0[\varphi^i]$ invariant under $m$ gauge transformations $\delta \varphi^i = R^i_\alpha \Lambda^\alpha$, which themselves are invariant under $n$ reducibility (`gauge-for-gauge') transformations $\delta \Lambda^\alpha = Z^\alpha_a \lambda^a$. This is equivalent to the identities
\begin{equation}
    \frac{\delta^R S_0}{\delta \varphi^i} R^i_\alpha = 0 \, , \quad R^i_\alpha Z^\alpha_a = 0 \, .
\end{equation}
We assume that there are no further reducibilities. The number of independent gauge redundancies in the fields $\varphi^i$ is therefore equal $m-n$. Accordingly, the gauge-fixing condition $\chi^\alpha(\varphi) = 0$ must only contain $m-n$ independent conditions. Since it carries a gauge index ranging from $1$ to $m$, we will take it to satisfy $n$ constraints:
\begin{equation}\label{eq:constrainedchi}
X_{a\alpha} \,\chi^\alpha(\varphi) = 0
\end{equation}
with $X_{a\alpha}$ of maximal rank.

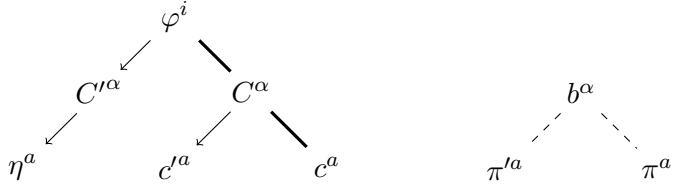
\begin{figure}
\centering
\begin{tikzpicture}
\node (ph) at (0,0) {$\varphi^i$};
\node (C0) at (1,-1) {$C^{\alpha}$};
\node (C1) at (2,-2) {$c^{a}$};
\node (C0p) at (-1,-1) {${C'}^{\alpha}$};
\node (C1p) at (-2,-2) {$\eta^{a}$};
\node (e) at (0,-2) {${c'}^{a}$};
\foreach \from/\to in {ph/C0, C0/C1}
	\draw [very thick] (\from) -- (\to);
\foreach \from/\to in {ph/C0p, C0p/C1p, C0/e}
	\draw [->] (\from) -- (\to);
\end{tikzpicture}
\qquad\qquad
\begin{tikzpicture}
\node (b) at (0,0) {$b^{\alpha}$};
\node (p1) at (1,-1) {$\pi^{a}$};
\node (p1p) at (-1,-1) {$\pi'^{a}$};
\foreach \from/\to in {b/p1, b/p1p}
	\draw [dashed] (\from) -- (\to);
\end{tikzpicture}
\caption{The pyramid of ghosts fields in the first-stage reducible case \cite{Batalin:1984jr}. The fields linked by a thick line constitute the minimal BV sector; an arrow $a \rightarrow b$ indicates that the field $b$ (along with its partner in a trivial pair) is introduced to fix the gauge freedom of $a$. The second pyramid shows the partners of the non-minimal fields of the first pyramid.}
\label{fig:pairsfirststage}
\end{figure}

In the minimal BV sector, we therefore have the original fields $\varphi^i$, the ghost $C^\alpha$ corresponding to the gauge parameter $\Lambda^\alpha$, and the ghost-for-ghost $c^a$ corresponding to the reducibility parameter $\lambda^a$, along with their antifields. The proper solution $\SM$ to the master equation starts as
\begin{equation}
\SM[\varphi^i, C^\alpha, c^a; \varphi^*_i, C^*_\alpha, c^*_a] = S_0[\varphi] + \varphi^*_i R^i_\alpha C^\alpha + C^*_\alpha Z^\alpha_a c^a + \dots \, .
\end{equation}
To build the non-minimal action, one introduces \emph{three} extra trivial pairs \cite{Batalin:1984jr}: $(C'^\alpha, b^\alpha)$ to fix the gauge freedom of $\varphi^i$, but also two more, $(c'^a, \pi^a)$ and $(\eta^a, \pi'^a)$, to fix the gauge freedom of the ghosts $C^\alpha$ and $C'^\alpha$ themselves. This is depicted in figure \ref{fig:pairsfirststage}. Their ghost numbers and Grassmann parities can be found in table \ref{tab:reducible}.
The non-minimal action is then
\begin{equation}
    \SNM = \SM + C'^*_\alpha b^\alpha + c'^*_a \pi^a + \eta^*_a \pi'^a\, .
\end{equation}

\begin{table}
    \centering
    \begin{tabular}{c| c c c c c c | c c c}
         & $\varphi_i$ & $C^\alpha$ & $C'^\alpha$ & $c^a$ & $c'^a$ & $\eta^a$ & $b^\alpha$ & $\pi^a$ & $\pi'^a$ \\ \midrule
        $\gh$ & $0$ & $1$ & $-1$ & $2$ & $-2$ & $0$ & $0$ & $-1$ & $1$ \\[\defaultaddspace]
        $\epsilon$ & $\epsilon_i$ & $\epsilon_\alpha+1$ & $\epsilon_\alpha+1$ & $\epsilon_a$ & $\epsilon_a$ & $\epsilon_a$ & $\epsilon_\alpha$ & $\epsilon_a+1$ & $\epsilon_a+1$
    \end{tabular}
    \caption{Ghost numbers and Grassmann parities of the various fields in the first-stage reducible case. Antifields have $\gh(\Phi^*_I) = - \gh(\Phi^I) - 1$ and opposite parity, $\epsilon(\Phi^*_I) = \epsilon(\Phi^I) + 1$.}
    \label{tab:reducible}
\end{table}

\paragraph{Delta function gauge-fixing.} This case is well-known \cite{Batalin:1984jr}: simply take the gauge-fixing fermion
\begin{equation}
    \Psi_\delta = C'_\alpha \chi^\alpha(\varphi) + c'^a \omega_{a\alpha} C^\alpha + \eta^a \sigma_{a}^{\alpha} C'_\alpha\, ,
\end{equation}
where $\omega$ and $\sigma$ are of maximal rank and we take the pair $(C'_\alpha, b_\alpha)$ to have indices down for this paragraph only. The gauge-fixed action then reads
\begin{align}
    S_\delta &\equiv \SNM\left[ \Phi^I, \Phi^*_I = \frac{\delta \Psi_\delta}{\delta \Phi^I} \right] \\
    &= \SM\left[ \varphi^i, C^\alpha, c^a; \varphi^*_i = C'_\alpha \frac{\delta^R \chi^\alpha}{\delta \varphi^i}, C^*_\alpha = c'^a \omega_{a\alpha}, c^*_a = 0 \right] \nn\\
    &\qquad + (\chi^\alpha(\varphi) + \eta^a \sigma_{a}^{\alpha})\, b_\alpha + \omega_{a\alpha} C^\alpha \pi^a + \sigma_{a}^{\alpha} C'_\alpha \pi'^a \, . \label{eq:reddelta}
\end{align}
In this action, the ghosts $C^\alpha$ and $C'_\alpha$ are both gauge fields. Their gauge invariances are fixed by the $2 m$ gauge conditions $\omega_{a\alpha} C^\alpha = 0$ and $\sigma_{a}^{\alpha} C'^\alpha = 0$ imposed by the auxiliary fields $\pi^a$ and $\pi'^a$. The field $b^\alpha$ is also auxiliary and imposes the equation
\begin{equation}\label{eq:chieta}
    \chi^\alpha(\varphi) + \eta^a \sigma_{a}^{\alpha} = 0\, .
\end{equation}
Among these $m$ conditions, $m-n$ fix the gauge invariance of the original fields $\varphi^i$, and the remaining $n$ set the extra ghost $\eta$ to zero.

\paragraph{Gaussian gauge-fixing.}

Now, we would like to achieve the gauge-fixing term $\cD_{\alpha\beta} \chi^\beta \chi^\alpha$. As before, the field $b^\alpha$ will become propagating if $\cD$ is a differential operator. However, an important difference with the irreducible case is that here $b^\alpha$ will be a constrained field, satisfying the same constraint \eqref{eq:constrainedchi} as the gauge condition.

One starts with the same canonical transformation \eqref{eq:canonicalirr} as in the irreducible case:
\begin{align}
    b^\alpha &\rightarrow \tilde{b}^\alpha = b^\alpha - \chi^\alpha(\varphi) \nn\\
    \varphi^*_i &\rightarrow \tilde{\varphi}^*_i = \varphi^*_i + b^*_\alpha \frac{\delta^R \chi^\alpha}{\delta \varphi^i}\label{eq:canonical}
\end{align}
with other fields unchanged. We take the gauge-fixing fermion
\begin{equation}
\Psi_\text{G} = \frac{1}{2} C'^\alpha \cD_{\alpha\beta}(\varphi) \left( \chi^\beta(\varphi) + b^\beta \right) + c'^a \omega_{a\alpha} C^\alpha + \eta^a \sigma_{a\alpha} C'^\alpha\, ,
\end{equation}
which is of the same form as $\Psi_\delta$, with only the first term modified along the lines of the irreducible case. Eliminating the antifields using $\Psi_G$ then gives
\begin{align}
    S_\text{G} &\equiv \tSNM\left[ \Phi^I, \Phi^*_I = \frac{\delta \Psi_\text{G}}{\delta \Phi^I} \right] \\
    &= \SM\left[ \varphi^i, C^\alpha, c^a; C'^\alpha \cD_{\alpha\beta} \frac{\delta^R \chi^\beta}{\delta \varphi^i} + \frac{1}{2} C'^\alpha \frac{\delta^R \cD_{\alpha\beta}}{\delta \varphi^i}\left( \chi^\beta + b^\beta \right) (-1)^{\epsilon_i \epsilon_\beta}, c'^a \omega_{\alpha a} ,  0 \right] \nonumber \\
    &\quad-\frac{1}{2} \cD_{\alpha\beta} \chi^\beta \chi^\alpha + \frac{1}{2}\cD_{\alpha\beta} b^\beta b^\alpha \\
    &\quad+ \eta^a \sigma_{a\alpha} (b^\alpha - \chi^\alpha) + \pi^a \omega_{a \alpha} C^\alpha  + \pi'^a \sigma_{a\alpha} C'^\alpha \, . \nonumber
\end{align}
Because of the constraint \eqref{eq:constrainedchi} satisfied by $\chi^\alpha(\varphi)$, there is a privileged choice for the matrix $\sigma_{a\alpha}$: simply take $\sigma = X$. This gets rid of the unwanted term $\eta^a \sigma_{a\alpha} \chi^\alpha$ in the last line, and one remains with
\begin{align}
    S_\text{G} = \SM&\left[ \varphi^i, C^\alpha, c^a; C'^\alpha \cD_{\alpha\beta} \frac{\delta^R \chi^\beta}{\delta \varphi^i} + \frac{1}{2} C'^\alpha \frac{\delta^R \cD_{\alpha\beta}}{\delta \varphi^i}\left( \chi^\beta + b^\beta \right) (-1)^{\epsilon_i \epsilon_\beta}, c'^a \omega_{\alpha a} ,  0 \right] \nn \\
    &-\frac{1}{2} \cD_{\alpha\beta} \chi^\beta \chi^\alpha + \frac{1}{2}\cD_{\alpha\beta} b^\beta b^\alpha + \pi^a \omega_{a \alpha} C^\alpha  + \pi'^a X_{a\alpha} C'^\alpha + \eta^a X_{a\alpha} b^\alpha\, , \label{eq:gaussianfirststage}
\end{align}
featuring the desired gauge-breaking term $\cD_{\alpha\beta} \chi^\beta \chi^\alpha$. Just as in the irreducible case, the field $b^\alpha$ is propagating whenever $\cD$ contains derivatives, and couples to the other fields and ghosts if $\cD$ is field-dependent. This generalises a result of \cite{Batalin:1983ar} to the reducible case.

In the action \eqref{eq:gaussianfirststage}, the auxiliary fields $\pi^a$ and $\pi'^a$ impose the gauge conditions
\begin{equation}
\omega_{a \alpha} C^\alpha = 0\, , \quad X_{a\alpha} C'^\alpha = 0
\end{equation}
on the ghost fields $C^\alpha$ and $C'^\alpha$, as in the delta-function gauge-fixing case. On the other hand, $\eta^a$ plays here a very different role as it did in \eqref{eq:reddelta}: it is now a Lagrange multiplier for the constraint
\begin{equation}
X_{a\alpha} \, b^\alpha = 0
\end{equation}
on the field $b^\alpha$. Notice how both $C'^\alpha$ and $b^\alpha$ satisfy the same constraint as $\chi^\alpha(\varphi)$ in this gauge-fixing scheme.

\subsection{Higher stage reducibility}\label{sec:BVredhigherstage}

This procedure generalises straightforwardly to theories with higher degree of reducibility. For concreteness, we write out the second-stage reducible case here. So, we consider an action $S_0[\varphi^i]$ with second-stage reducible gauge symmetries:
\begin{equation}
    \delta\varphi^i = R^i{}_{\alpha_0} \Lambda^{\alpha_0}\, , \quad \delta\Lambda^{\alpha_0} = Z^{\alpha_0}{}_{\alpha_1} \lambda^{\alpha_1}\, , \quad \delta \lambda^{\alpha_1} = z^{\alpha_1}{}_{\alpha_2} \epsilon^{\alpha_2}
\end{equation}
with $\alpha_0 = 1, \dots, m$, $\alpha_1 = 1, \dots, n$ and $\alpha_2 = 1, \dots, r$. Invariance under these transformations is equivalent to the relations
\begin{equation}
    \frac{\delta^R S_0}{\delta \varphi^i} R^i{}_{\alpha_0} = 0\, , \quad R^i{}_{\alpha_0}Z^{\alpha_0}{}_{\alpha_1} = 0\, , \quad Z^{\alpha_0}{}_{\alpha_1}z^{\alpha_1}{}_{\alpha_2} = 0\, ,
\end{equation}
and we assume that there are no further reducibilities.
The gauge condition $\chi^{\alpha_0}(\varphi) = 0$ must fix the $m-n+r$ independent gauge transformations: we take it to satisfy constraints $X_{\alpha_1 \alpha_0} \chi^{\alpha_0} = 0$ as in the previous case, but here with a degenerate matrix $X$ of rank $n-r$.

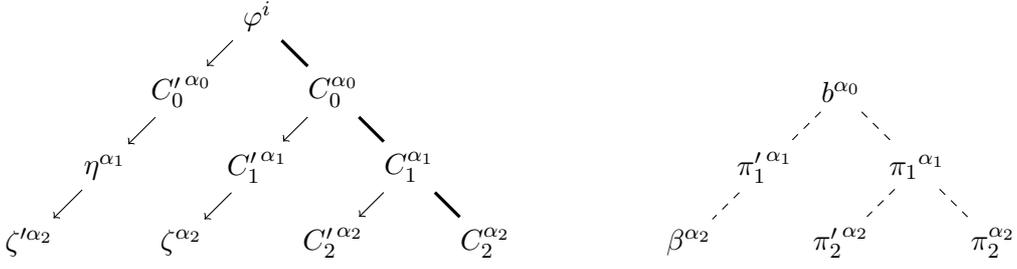
\begin{figure}
\centering
\begin{tikzpicture}
\node (ph) at (0,0) {$\varphi^i$};
\node (C0) at (1,-1) {$C_0^{\alpha_0}$};
\node (C1) at (2,-2) {$C_1^{\alpha_1}$};
\node (C2) at (3,-3) {$C_2^{\alpha_2}$};
\node (C0p) at (-1,-1) {${C_0'}^{\alpha_0}$};
\node (C1p) at (-2,-2) {$\eta^{\alpha_1}$};
\node (C2p) at (-3,-3) {$\zeta'^{\alpha_2}$};
\node (e) at (0,-2) {${C_1'}^{\alpha_1}$};
\node (z) at (1,-3) {${C_2'}^{\alpha_2}$};
\node (zp) at (-1,-3) {$\zeta^{\alpha_2}$};
\foreach \from/\to in {ph/C0, C0/C1, C1/C2}
	\draw [very thick] (\from) -- (\to);
\foreach \from/\to in {ph/C0p, C0p/C1p, C1p/C2p, C0/e, e/zp, C1/z}
	\draw [->] (\from) -- (\to);
\end{tikzpicture}
$\qquad\qquad$
\begin{tikzpicture}
\node (b) at (0,0) {$b^{\alpha_0}$};
\node (p1) at (1,-1) {${\pi_1}^{\alpha_1}$};
\node (p2) at (2,-2) {$\pi_2^{\alpha_2}$};
\node (p1p) at (-1,-1) {${\pi_1'}^{\alpha_1}$};
\node (p2p) at (-2,-2) {$\beta^{\alpha_2}$};
\node (be) at (0,-2) {${\pi_2'}^{\alpha_2}$};
\foreach \from/\to in {b/p1, p1/p2, b/p1p, p1p/p2p, p1/be}
	\draw [dashed] (\from) -- (\to);
\end{tikzpicture}
\caption{The pyramids of ghost fields in the second-stage reducible case \cite{Batalin:1984jr}. This should be read in the same way as figure \ref{fig:pairsfirststage}.}
\label{fig:pairssecondstage}
\end{figure}

In the minimal BV sector, there are now three generations of ghosts: $C_0^{\alpha_0}$, $C_1^{\alpha_1}$ and $C_2^{\alpha_2}$. The first few terms in the minimal action are simply
\begin{align}
    &S^M[\varphi^i,C_0^{\alpha_0},C_1^{\alpha_1},C_2^{\alpha_2};\,\varphi^*_i,{C_0^*}_{\alpha_0},{C_1^*}_{\alpha_1},{C_2^*}_{\alpha_2}] \\
    &\qquad = S_0[\varphi^i] + \varphi^*_i R^i{}_{\alpha_0} C_0^{\alpha_0} + {C_0^*}_{\alpha_0} Z^{\alpha_0}{}_{\alpha_1} C_1^{\alpha_1} + {C_1^*}_{\alpha_1} z^{\alpha_1}{}_{\alpha_2} C_2^{\alpha_2} + \dots\, .\nn
\end{align}
For gauge-fixing, we add the usual extra pairs, as in figure \ref{fig:pairssecondstage}.
The non-minimal action, after the canonical transformation \eqref{eq:canonical}, then reads
\begin{align}
    \tSNM &= \SM\Big[\varphi^i,C_0^{\alpha_0},C_1^{\alpha_1},C_2^{\alpha_2};\,\varphi^*_i + b^*_{\alpha_0} \frac{\delta^R \chi^{\alpha_0}}{\delta \varphi^i},{C_0^*}_{\alpha_0},{C_1^*}_{\alpha_1},{C_2^*}_{\alpha_2}\Big] \\
    &\quad + {C_0'}^*_{\alpha_0} \left( b^{\alpha_0} - \chi^{\alpha_0}(\varphi) \right) + {C_1'}^*_{\alpha_1} {\pi_1}^{\alpha_1} + \eta^*_{\alpha_1} {\pi'_1}^{\alpha_1} \nn \\
    &\quad+ {C_2'}^*_{\alpha_2} {\pi_2}^{\alpha_2} + {\zeta}^*_{\alpha_2} {\pi'_2}^{\alpha_2} + {\zeta'}^*_{\alpha_2} \beta^{\alpha_2} \, . \nn
\end{align}
We use the gauge-fixing fermion
\begin{align}
    \Psi_\text{G} &= \frac{1}{2}{C_0'}^{\alpha_0} \cD_{\alpha_0 \beta_0}(\varphi) (\chi^{\beta_0}(\varphi) + b^{\beta_0}) + {C_1'}^{\alpha_1} (\omega_1)_{\alpha_1 \alpha_0} {C_0}^{\alpha_0} + {C_2'}^{\alpha_2} (\omega_2)_{\alpha_2 \alpha_1} {C_1}^{\alpha_1} \nn\\
    &\quad + \eta^{\alpha_1} X_{\alpha_1 \alpha_0} {C_0'}^{\alpha_0} + \zeta^{\alpha_2} (\sigma_2)_{\alpha_2 \alpha_1} {C_1'}^{\alpha_1} + \zeta'^{\alpha_2} (\sigma'_2)_{\alpha_2 \alpha_1} \eta^{\alpha_1}\, ,
\end{align}
where $\omega_1$ and $X$ are of rank $n-r$ and $\omega_2$, $\sigma_2$ and $\sigma'_2$ are of maximal rank $r$. The gauge-fixed action is then
\begin{align}
    S_\text{G} &= \SM\Big[\varphi^i,C_0^{\alpha_0},C_1^{\alpha_1},C_2^{\alpha_2};\nn\\
    &\qquad \qquad\,{C_0'}^{\alpha_0} \cD_{\alpha_0\beta_0} \frac{\delta^R \chi^{\beta_0}}{\delta \varphi^i} + \frac{1}{2} {C_0'}^{\alpha_0} \frac{\delta^R \cD_{\alpha_0\beta_0}}{\delta \varphi^i}\left( \chi^{\beta_0} + b^{\beta_0} \right) (-1)^{\epsilon_i \epsilon_{\beta_0}},\nn \\
    &\qquad \qquad\,{C_1'}^{\alpha_1} (\omega_1)_{\alpha_1 \alpha_0},{C_2'}^{\alpha_2} (\omega_2)_{\alpha_2 \alpha_1},0\Big] \nn \\
    &\quad -\frac{1}{2} \cD_{\alpha_0\beta_0} \chi^{\beta_0} \chi^{\alpha_0} + \frac{1}{2}\cD_{\alpha_0\beta_0} b^{\beta_0} b^{\alpha_0} + \eta^{\alpha_1} (X_{\alpha_1 \alpha_0} b^{\alpha_0} + (\sigma'_2)_{\alpha_2 \alpha_1} \beta^{\alpha_2})\nn \\
    &\quad + ((\omega_1)_{\alpha_1 \alpha_0} {C_0}^{\alpha_0} + \zeta^{\alpha_2} (\sigma_2)_{\alpha_2 \alpha_1} ){\pi_1}^{\alpha_1} + (X_{\alpha_1 \alpha_0} {C_0'}^{\alpha_0} + \zeta'^{\alpha_2} (\sigma'_2)_{\alpha_2 \alpha_1}) {\pi'_1}^{\alpha_1} \nn \\
    &\quad+ (\omega_2)_{\alpha_2 \alpha_1} {C_1}^{\alpha_1} {\pi_2}^{\alpha_2} + (\sigma_2)_{\alpha_2 \alpha_1} {C_1'}^{\alpha_1} {\pi'_2}^{\alpha_2}\, ,
\end{align}
with the same structure as \eqref{eq:gaussianfirststage} and where we already used the constraint $X_{\alpha_1 \alpha_0} \chi^{\alpha_0} = 0$. The auxiliary fields $\eta^{\alpha_1}$, ${\pi_1}^{\alpha_1}$ and ${\pi_1'}^{\alpha_1}$ impose the constraints
\begin{align}
X_{\alpha_1 \alpha_0} b^{\alpha_0} + (\sigma'_2)_{\alpha_2 \alpha_1} \beta^{\alpha_2} &= 0 \\
(\omega_1)_{\alpha_1 \alpha_0} {C_0}^{\alpha_0} + (\sigma_2)_{\alpha_2 \alpha_1} \zeta^{\alpha_2} &= 0 \\
X_{\alpha_1 \alpha_0} {C_0'}^{\alpha_0} + (\sigma'_2)_{\alpha_2 \alpha_1} \zeta'^{\alpha_2} &= 0
\end{align}
which give $n-r$ constraints on $b^{\alpha_0}$, ${C_0}^{\alpha_0}$, ${C_0'}^{\alpha_0}$ and imply the vanishing of the extra ghosts, $\beta^{\alpha_2} = \zeta^{\alpha_2} = \zeta'^{\alpha_2} = 0$. The fields ${\pi_2}^{\alpha_2}$ and ${\pi_2'}^{\alpha_2}$ impose the gauge-fixing conditions
\begin{equation}
(\omega_2)_{\alpha_2 \alpha_1} {C_1}^{\alpha_1} = 0\, , \quad (\sigma_2)_{\alpha_2 \alpha_1} {C_1'}^{\alpha_1} = 0\, .
\end{equation}

\section{Free fermionic \texorpdfstring{$p$}{}-form fields}\label{sec:pforms}

In this section, we apply the BV formalism to the quantisation of antisymmetric tensor spinors. Since fermionic fields satisfy first-order equations of motion and the action \eqref{eq:actionintro} is already in Hamiltonian form, the Hamiltonian quantisation methods of \cite{Fradkin:1975cq,Batalin:1977pb,Fradkin:1977xi} would have been more economical. The third ghost has also been discussed in that formalism in reference \cite{Henneaux:1983bz}. However, the approach we use here has the advantage of preserving manifest covariance.

\subsection{Action, gauge symmetries and gauge conditions}\label{sec:pformactionetc}

The action for a fermionic $p$-form field, that is, a tensor-spinor $\psi^\alpha_{\mu_1\mu_2 \dots \mu_p}$ totally antisymmetric in its spacetime indices, was already presented in the introduction: it is given by \cite{Buchbinder:2009pa, Zinoviev:2009wh,Campoleoni:2009gs}
\begin{equation}\label{eq:S0-pform}
    S_0[\psi] = - (-1)^{\frac{p(p-1)}{2}} \, \int \! d^D\!x\, \bar{\psi}_{\mu_1\mu_2\dots \mu_p} \,\gamma^{\mu_1\mu_2\dots \mu_p \nu \rho_1\rho_2\dots \rho_p} \,\pd_\nu \psi_{\rho_1\rho_2\dots \rho_p}\, .
\end{equation}
Due to the rank $2p+1$ antisymmetric gamma matrix, it is manifestly invariant under the gauge symmetries
\begin{equation}\label{eq:gaugepform}
\delta \psi^\alpha_{\mu_1\mu_2\dots \mu_p} = p \, \pd_{[\mu_1} {\Lambda^{(p-1)}}{}^{\alpha}_{\mu_2\dots \mu_p]}\, ,
\end{equation}
where the gauge parameter $\Lambda^{(p-1)}$ is an arbitrary antisymmetric tensor-spinor of rank $p-1$. This system is $(p-1)$-stage reducible: \eqref{eq:gaugepform} comes with the chain of gauge-for-gauge transformations
\begin{align}
\delta {\Lambda^{(p-1)}}{}^{\alpha}_{\mu_2\dots \mu_p} &= (p-1)\,\pd_{[\mu_2}{\Lambda^{(p-2)}}{}^{\alpha}_{\mu_3\dots \mu_p]} \\
\delta {\Lambda^{(p-2)}}{}^{\alpha}_{\mu_3\dots \mu_p} &= (p-2)\, \pd_{[\mu_3}{\Lambda^{(p-3)}}{}^{\alpha}_{\mu_4\dots \mu_p]} \\
&\;\;\vdots \nn \\
\delta {\Lambda^{(1)}}{}^{\alpha}_{\mu} &= \pd_{\mu}{\Lambda^{(0)}}{}^{\alpha}\, ,
\end{align}
where each parameter $\Lambda^{(k)}$ is a rank-$k$ antisymmetric tensor-spinor. In differential form notation with a spectator spinor index, this is
\begin{equation}\label{eq:pformgauge}
\delta \psi = d \Lambda^{(p-1)}\, , \quad \delta \Lambda^{(p-1)} = d\Lambda^{(p-2)}\, , \quad \dots \, , \quad \delta \Lambda^{(1)} = d\Lambda^{(0)}\, .
\end{equation}

The equations of motion coming from the action \eqref{eq:S0-pform} read
\begin{equation}
\gamma^{\mu_1 \dots \mu_p \nu_1 \dots \nu_{p+1}} \,H_{\nu_1 \dots \nu_{p+1}} = 0\, , \quad H_{\mu_1 \dots \mu_{p+1}} \equiv (p+1)\, \pd_{[\mu_1} \psi_{\mu_2 \dots \mu_{p+1}]}\, ,
\end{equation}
where $H = d\psi$ is the gauge-invariant field strength of the field $\psi$. Equivalently, they can be written as the single-gamma-trace equation
\begin{equation}
\gamma^{\mu_1} H_{\mu_1 \mu_2 \dots \mu_{p+1}} = 0\, .
\end{equation}
These equations propagate the correct representation of the massless little group: the rank $p$ antisymmetric tensor-spinor of $SO(D-2)$ satisfying a gamma-tracelessness condition. Such a tensor-spinor identically vanishes for $p \geq D/2$. This is consistent with the covariant action \eqref{eq:S0-pform}: that action identically vanishes when $2p+1 > D$ because of the antisymmetric gamma matrix and, for $2p + 1 = D$, the equations of motion are equivalent to $H[\psi] = 0$ which implies that $\psi$ is pure gauge. So, the theory described by \eqref{eq:S0-pform} has propagating degrees of freedom only for $2p < D$, and we will assume this inequality for the remainder of this section.

We now turn to the gauge condition that we will impose on the field $\psi^\alpha_{\mu_1\mu_2\dots \mu_p}$. It is given by an equation of the form
\begin{equation}
\chi^\alpha_{\mu_1 \dots \mu_{p-1}}(\psi) = 0\, ,
\end{equation}
with the same index structure as the gauge parameter, that must only contain as many independent conditions as there are independent gauge transformations. That number is
\begin{equation}\label{eq:gaugecounting}
N_{p-1} - N_{p-2} + N_{p-3} - N_{p-4} + \cdots \pm N_0
\end{equation}
where $N_k$ is the number of components of an antisymmetric tensor-spinor of rank $k$ and where the final sign depends on the parity of $p$ (the precise formula for $N_k$ is irrelevant for the sake of the argument). One way to realise this is to take a gauge condition $\chi^\alpha_{\mu_1 \dots \mu_{p-1}}(\psi)$ that satisfies $N_{p-2} - N_{p-3} + N_{p-4} - \cdots$ independent constraints; this can work if the operator $X$ in the constraint equations $X^\alpha_{\mu_1 \dots \mu_{p-2}}(\chi) = 0$ (cf.~section \ref{sec:BVred}) itself satisfies $N_{p-3} - N_{p-4} + \cdots$ independent constraints, etc. This reasoning shows that it is sufficient to define operators $T^{(k)}$ mapping fermionic $k$-forms to $(k-1)$-forms such that the nilpotency condition
\begin{equation}
    T^{(k)} \circ T^{(k+1)} = 0
\end{equation}
holds and exhausts the constraints satisfied by $T^{(k+1)}$ (extra constraints would of course upset the counting above). Then, the gauge condition
\begin{equation}
\chi(\psi) \equiv T^{(p)}(\psi) = 0
\end{equation}
satisfies $T^{(p-1)}(\chi) = 0$ and gives the correct number of independent conditions. From the analysis of the Rarita-Schwinger case ($p=1$), eq.~\eqref{eq:gaugecondRS}, we take
\begin{equation}
T^{(1)}(\psi) = \gamma^\mu \psi_\mu    
\end{equation}
as a suitable starting point. The next operators can then be determined recursively using the nilpotency condition: the first few read explicitly
\begin{align}
    T^{(2)}(\psi)_\mu &= \gamma^\nu \psi_{\mu\nu} - \frac{1}{D} \gamma_\mu \gamma^{\nu\rho} \psi_{\nu\rho} \\
    T^{(3)}(\psi)_{\mu\nu} &= \gamma^\rho \psi_{\mu\nu\rho} + \frac{2}{D-2} \gamma_{[\mu} \gamma^{\rho\sigma} \psi_{\nu]\rho\sigma} \\
    T^{(4)}(\psi)_{\mu\nu\rho} &= \gamma^\sigma \psi_{\mu\nu\rho\sigma} - \frac{3}{D-4} \gamma_{[\mu} \gamma^{\sigma\tau} \psi_{\nu\rho]\sigma\tau} - \frac{2}{D(D-2)(D-4)}\gamma_{\mu\nu\rho} \gamma^{\sigma\tau\kappa\lambda} \psi_{\sigma\tau\kappa\lambda} \\
     T^{(5)}(\psi)_{\mu\nu\rho\sigma} &= \gamma^\tau \psi_{\mu\nu\rho\sigma\tau} + \frac{4}{D-6} \gamma_{[\mu} \gamma^{\tau\kappa} \psi_{\nu\rho\sigma]\tau\kappa} + \frac{8}{(D-2)(D-4)(D-6)} \gamma_{[\mu\nu\rho} \gamma^{\tau\kappa\lambda\zeta}  \psi_{\sigma]\tau\kappa\lambda\zeta}\, .
\end{align}

\subsection{Quantisation of the fermionic \texorpdfstring{$2$}{}-form}

In this section, we carry out the quantisation of the fermionic $2$-form explicitly along the lines explained in section \ref{sec:BVredfirststage}, both in delta-function gauge-fixing and in the Gaussian gauge-fixing where an extra Nielsen-Kallosh ghost appears. The generalisation to higher form degrees is direct.

The action, gauge transformations and reducibilities read
\begin{equation}\label{eq:S0-twoform}
    S_0[\psi] = \int \! d^D\!x\, \bar{\psi}_{\mu\nu} \gamma^{\mu\nu\rho\sigma\tau} \pd_\rho \psi_{\sigma\tau}\, , \quad \delta \psi^\alpha_{\mu\nu} = 2\, \pd_{[\mu} \Lambda^\alpha_{\nu]}\, , \quad \delta \Lambda^\alpha_\mu = \pd_\mu \lambda^\alpha\, .
\end{equation}
The gauge parameter $\Lambda$ has $n=Ds$ components and $\lambda$ has $m=s$ components, for a total of $n-m=s(D-1)$ independent gauge symmetries, where $s$ is the dimension of the spinor representation at hand. (In this section as in section \ref{sec:RSquant}, we consider Dirac spinors so $s=2^{[D/2]}$, but this counting is of course also valid with when reality and/or chirality conditions are imposed on the fields.) This action was used in the papers \cite{Henneaux:2017xsb,Henneaux:2018rub} as part of a complete free action principle for the exotic $\cN = (4,0)$ and $\cN = (3,1)$ multiplets in $D=6$.

Accordingly, the minimal spectrum in the Batalin-Vilkovisky formalism consists of the fields and antifields
\begin{equation}
    \{ \psi^\alpha_{\mu\nu}\, , C^\alpha_\mu\, , c^\alpha\, , \psi_\alpha^{*\mu\nu}\, , C^{*\mu}_\alpha\, , c^*_\alpha \}\, ,
\end{equation}
where $C^\alpha_\mu$ is the ghost associated to the $\Lambda^\alpha_\mu$ gauge parameter and $c^\alpha$ is the ghost-for-ghost associated to the reducibility parameter $\lambda^\alpha$. The minimal BV master action reads
\begin{align}
    \SM &= \int \! d^D\!x\, \left( \frac{1}{2}\bar{\psi}_{\mu\nu} \gamma^{\mu\nu\rho\sigma\tau} \pd_\rho \psi_{\sigma\tau} + 2\, \psi^{*\mu\nu} \pd_{\mu} C_\nu + C^{*\mu} \pd_\mu c +\cc\right)\, .
\end{align}
Note that antifields transform naturally as a conjugate spinor. The non-minimal action, with the usual trivial pairs, is
\begin{align}\label{eq:SNM-2form}
    \SNM &= \SM + \int \! d^D\!x\, \left( C'^{*\mu} b_\mu + c'^* \pi + \eta^* \pi' + \cc \right)\, .
\end{align}
Both are ghost number zero functionals satisfying the master equation
\begin{equation}
(\SM,\SM) = 0 = (\SNM,\SNM)\, .
\end{equation}
We will use the redundant gauge condition
\begin{align}\label{eq:gaugecond2form}
    \chi_\mu(\psi) &\equiv \gamma^\nu \psi_{\mu\nu} - \frac{1}{D-2} \gamma_{\mu\nu\rho} \psi^{\nu\rho} \nn \\
    &= 0 \, ,
\end{align}
which satisfies the constraint
\begin{equation}
    \gamma^\mu \chi_\mu(\psi) = 0
\end{equation}
identically and hence gives the correct number $n-m=s(D-1)$ of gauge conditions to fix the independent gauge transformations. In the notation of the previous section, this is $\chi = \frac{D}{D-2} T^{(1)}$, with a convenient rescaling. To understand this gauge condition better, it is useful to write the different trace components of $\psi_{\mu\nu}$ explicitly:
\begin{equation}\label{eq:decomppsi}
    \psi_{\mu\nu} = \hat{\psi}_{\mu\nu} + ( \gamma_\mu \sigma_\nu - \gamma_\nu \sigma_\mu ) + \gamma_{\mu\nu} \rho
\end{equation}
where $\hat{\psi}_{\mu\nu}$ and $\sigma_\mu$ are gamma-traceless, $\gamma^\nu \hat{\psi}_{\mu\nu} = 0 = \gamma^\mu \sigma_\mu$. A short computation then shows that the condition $\chi_\mu(\psi) = 0$ is equivalent to $\sigma_\mu = 0$, i.e.~setting the spin $3/2$ component $\sigma_\mu$ to zero but \emph{not} the spin $1/2$ part $\rho$, indeed removing $s(D-1)$ components of $\psi_{\mu\nu}$. The same is valid more generally for $p$-forms with the gauge conditions of the previous section: $T^{(p)}(\psi) = 0$ kills all gamma-traceless components of rank $p-1$, $p-3$, etc. This is consistent with the counting \eqref{eq:gaugecounting} of independent gauge transformations.

\paragraph{Delta-function gauge-fixing.}

The gauge-fixing fermion is taken as
\begin{align}\label{eq:psidelta}
    \Psi_\delta &= \int \! d^D\!x\, \left( \bar{C}'_\mu\, \chi^\mu(\psi) + \bar{c}' \gamma^\mu C_\mu + \bar{\eta} \gamma^\mu C'_\mu  +\cc \right)\, ,
\end{align}
leading to
\begin{align}\label{eq:Sgf-twoform-delta}
    S_\delta = \int \! d^D\!x\, \Big( &\frac{1}{2}\bar{\psi}_{\mu\nu} \gamma^{\mu\nu\rho\sigma\tau} \pd_\rho \psi_{\sigma\tau} + 2\, \bar{C}'^{\sigma} \frac{\delta \chi_\sigma}{\delta \psi_{\mu\nu}} \pd_\mu C_\nu + \bar{c}' \slashed{\pd} c \nn \\
    &+ \bar{b}_\mu \left( \chi^{\mu}(\psi) - \gamma^\mu \eta \right) + \bar{\pi} \gamma^\mu C_\mu + \bar{\pi}' \gamma^\mu C'_\mu + \cc \Big)
\end{align}
where
\begin{equation}
\frac{\delta \chi_\sigma}{\delta \psi_{\mu\nu}} = \delta^{[\mu}_\sigma \gamma^{\nu]} - \frac{1}{D-2} \gamma_\sigma{}^{\mu\nu}\, .
\end{equation}
The auxiliary fields enforce the gauge conditions
\begin{equation}
    \chi^{\mu}(\psi) - \gamma^\mu \eta = 0\, , \quad \gamma^\mu C_\mu = 0\, , \quad \gamma^\mu C'_\mu = 0\, .
\end{equation}
Contracting the first condition with $\gamma_\mu$ gives $\eta = 0$ owing to the constraint satisfied by $\chi^{\mu}(\psi)$, and then it also implies the gauge condition $\chi^{\mu}(\psi) = 0$. 

Using this gauge condition, the kinetic term for $\psi$ can be simplified using the decomposition \eqref{eq:decomppsi} with $\sigma_\mu = 0$, which gives
\begin{equation}
    \frac{1}{2}\bar{\psi}_{\mu\nu} \gamma^{\mu\nu\rho\sigma\tau} \pd_\rho \psi_{\sigma\tau} = - \bar{\hat{\psi}}^{\mu\nu} \slashed{\pd} \hat{\psi}_{\mu\nu} - \frac{1}{2}(D-1)(D-2)(D-3)(D-4)\, \bar{\rho} \slashed{\pd} \rho\, .
\end{equation}
The gamma-tracelessness conditions on $C_\mu$ and $C'_\mu$ can also be used to reduce their kinetic term to
\begin{equation}
    2 \,\bar{C}'^{\sigma} \frac{\delta \chi_\sigma}{\delta \psi_{\mu\nu}} \pd_\mu C_\nu = - \frac{2 D}{D-2} \bar{C}'^{\mu} \slashed{\pd} C_\mu\, .
\end{equation}
Rescaling the fields and using an auxiliary field $d^\mu$ to impose the gamma-tracelessness of $\hat{\psi}_{\mu\nu}$, the final result can be written as 
\begin{align}
    S_\delta = \int \! d^D\!x\, \Big( &- \frac{1}{2}\bar{\hat{\psi}}^{\mu\nu} \slashed{\pd} \hat{\psi}_{\mu\nu} - \frac{1}{2}\bar{\rho} \slashed{\pd} \rho + \bar{C}'^{\mu} \slashed{\pd} C_\mu + \bar{c}' \slashed{\pd} c \nn \\
    &+ \bar{d}^\mu \gamma^\nu \hat{\psi}_{\mu\nu} + \bar{\pi} \gamma^\mu C_\mu + \bar{\pi}' \gamma^\mu C'_\mu + \cc \Big)\, .\label{eq:Sgf-twoform-delta-final}
\end{align}
The spectrum of dynamical fields is as follows: one gamma-traceless $2$-form, two gamma-traceless one-forms and three zero-forms, with alternating Grassmann parity (hence spin-statistics) at every stage. In terms of gamma-traceful fields, this formally corresponds to $1$, $3$ and $5$ fields respectively. The same pattern appears for higher degree: one gets $1$, $2$, $3$, $4$, ... gamma-traceless forms of descending degree and alternating parity, which effectively corresponds to $1$, $3$, $5$, $7$, ... gamma-traceful fields. This is as expected in a reducible fermionic theory \cite{Siegel:1980jj}.

\paragraph{Gaussian gauge-fixing.} We now would like to achieve a gauge-fixing term of the form $\bar{\chi}_\mu \cD^{\mu\nu} \chi_\nu$, where $\cD^{\mu\nu}$ is some first-order differential operator. Using only gamma matrices, the flat metric $\eta^{\mu\nu}$ and one space-time derivative, we find using the gamma-tracelessness of $\chi_\mu$ that the only independent possibility for $\cD^{\mu\nu}$ is the very simple
\begin{equation}
\cD^{\mu\nu} = \eta^{\mu\nu} \slashed{\pd}\, .
\end{equation}
As indicated in section \ref{sec:BVredfirststage}, we start with the canonical transformation
\begin{align}
    b_\mu &\rightarrow b_\mu - \chi_\mu(\psi) \, , \quad \psi^{*\mu\nu} \rightarrow \psi^{*\mu\nu} + b^{*\sigma} \frac{\delta \chi_\sigma}{\delta \psi_{\mu\nu}} \, ,
\end{align}
which gives the new non-minimal action
\begin{align}\label{eq:SNMtilde-2form}
    \tSNM = \int \! d^D\!x\, \Big( \frac{1}{2}\bar{\psi}_{\mu\nu} \gamma^{\mu\nu\rho\sigma\tau} \pd_\rho \psi_{\sigma\tau} &+ 2 \left( \psi^{*\mu\nu} + b^{*\sigma} \frac{\delta \chi_\sigma}{\delta \psi_{\mu\nu}} \right) \pd_{\mu} C_\nu + C^{*\mu} \pd_\mu c \nn \\
    &+ C'^{*\mu} (b_\mu - \chi_\mu) + c'^* \pi + \eta^* \pi' + \cc \Big)\, ,
\end{align}
and we use the gauge-fixing fermion
\begin{align}\label{eq:psiG}
    \Psi_\text{G} &= \int \! d^D\!x\, \left( -\frac{\xi}{2}\bar{C}'^\sigma \slashed{\pd}\, (b_\sigma + \chi_\sigma)+ \bar{c}' \gamma^\mu C_\mu + \bar{\eta} \gamma^\mu C'_\mu  +\cc \right)
\end{align}
where $\xi \neq 0$ is an arbitrary parameter.
This gives the gauge-fixed action
\begin{align}\label{eq:Sgftwoform1}
    S_\text{G} = \int \! d^D\!x\, \Big( &\frac{1}{2}\bar{\psi}_{\mu\nu} \gamma^{\mu\nu\rho\sigma\tau} \pd_\rho \psi_{\sigma\tau} -2 \xi\, \bar{C}'^\sigma \, \slashed{\pd}\,\frac{\delta \chi_\sigma}{\delta \psi_{\mu\nu}} \pd_\mu C_\nu + \bar{c}' \slashed{\pd} c -\frac{\xi}{2} \bar{b}_\mu \,\slashed{\pd} \,b^\mu  \nn \\
    &+ \frac{\xi}{2} \bar{\chi}_\mu \,\slashed{\pd} \, \chi^\mu + \bar{\eta} \gamma^\mu (b_\mu + \chi_\mu) + \bar{\pi} \gamma^\mu C_\mu + \bar{\pi}' \gamma^\mu C'_\mu + \cc \Big)\, .
\end{align}
The term $\bar{\eta} \gamma^\mu \chi_\mu(\psi)$ in this action identically vanishes thanks to the constraint satisfied by $\chi_\mu$. The auxiliary fields $\eta$, $\pi$ and $\pi'$ impose the gamma-tracelessness of $C_\mu$, $C'_\mu$ and $b_\mu$.

Using the gamma-tracelessness conditions on the ghosts $C_\mu$ and $C'_\mu$, their kinetic term can be simplified as
\begin{align}
    -2\xi\, \bar{C}'^\sigma \, \slashed{\pd}\,\frac{\delta \chi_\sigma}{\delta \psi_{\mu\nu}} \pd_\mu C_\nu &= \frac{2\xi D}{D-2} \bar{C}'_{\mu} \Box C^\mu
\end{align}
The final result is then, after rescaling some of the fields,
\begin{align}\label{eq:Sgf-twoform-gaussian}
    S_\text{G} = \int \! d^D\!x\, \bigg[ &\bar{\psi}_{\mu\nu} \gamma^{\mu\nu\rho\sigma\tau} \pd_\rho \psi_{\sigma\tau} + \xi\, \bar{\chi}_\mu(\psi) \,\slashed{\pd}\, \chi^\mu(\psi) \\
    &+ ( \bar{C}'_{\mu} \Box C^\mu + \bar{C}_{\mu} \Box C'^\mu ) -\, \bar{b}_\mu \,\slashed{\pd} \,b^\mu + (\bar{c}' \slashed{\pd} c + \bar{c} \slashed{\pd} c') \nn \\
    &+ (\bar{\eta} \gamma^\mu b_\mu - \bar{b}_\mu \gamma^\mu \eta) + (\bar{\pi} \gamma^\mu C_\mu - \bar{C}_\mu \gamma^\mu \pi) + (\bar{\pi}' \gamma^\mu C'_\mu - \bar{C}'_\mu \gamma^\mu \pi') \bigg] \nn
\end{align}
with the desired gauge-breaking term. The field $b_\mu$ has a kinetic term and is a propagating spin $3/2$ field: it is the Nielsen-Kallosh ghost for the fermionic two-form.
As in the Rarita-Schwinger case, the ghosts $C_\mu$ and $C'_\mu$ have a second-order kinetic term and hence count for four. The field $b_\mu$ has the correct spin-statistics and there are effectively three spin $3/2$ ghosts, as expected.

\paragraph{BRST transformations.}

The gauge-fixed actions above are invariant under a nilpotent BRST transformation of ghost number $+1$, and the extra terms in the action (gauge-breaking terms and ghosts terms) are BRST-exact. This comes very naturally out of the field-antifield formalism; we refer to the reviews \cite{Henneaux:1992ig,Gomis:1994he} for a general discussion.

In the delta-function gauge-fixing case, the action of the BRST differential $\bar{s}$ on a functional $A$ depending on the fields $\Phi^I$ of the non-minimal sector (but not on the antifields $\Phi^*_I$) is given by
\begin{equation}\label{eq:sdelta}
    \bar{s} A = (A, \SNM)\Big\vert_{\Phi^* = \tfrac{\delta \Psi_\delta}{\delta \Phi}}\, = \left.\frac{\delta^R A}{\delta \Phi^I}\frac{\delta^L \SNM}{\delta \Phi^*_I}\right|_{\Phi^* = \tfrac{\delta \Psi_\delta}{\delta \Phi}}\, ,
\end{equation}
where the non-minimal action is in eq.~\eqref{eq:SNM-2form}. Notice, however, that \eqref{eq:SNM-2form} is linear in the antifields\footnote{Terms of higher order in antifields would be expected in a putative interacting theory with a more involved gauge structure, e.g.~if the gauge algebra were open.}: therefore, $\frac{\delta^L S^{NM}}{\delta \Phi^*_I}$ is antifield-independent and the definition of $s$ is in fact does not depend on the gauge-fixing fermion. On the fields, $\bar{s}$ explicitly reads
\begin{align}\label{eq:sbar}
    \bar{s}\psi_{\mu\nu} &= 2 \pd_{[\mu} C_{\nu]}\, , \quad \bar{s} C_\mu = \pd_\mu c\, , \quad \bar{s} C'_\mu = b_\mu\, , \quad \bar{s} c' = \pi \, , \quad \bar{s} \eta = \pi' \, , \quad \bar{s} (\text{other}) = 0 \, .
\end{align}
The nilpotency 
\begin{equation}
\bar{s}\,{}^2 = 0
\end{equation}
is immediate and holds off-shell. On $\psi_{\mu\nu}$ (and $C_\mu$ due to the reducibility), $\bar{s}$ takes of course the familiar form `gauge transformations with parameter replaced by ghost'. The gauge-fixed action \eqref{eq:Sgf-twoform-delta} can then be written as
\begin{equation}
    S_{\delta} = S_0 + \bar{s}\, \Psi_\delta\, ,
\end{equation}
with $S_0$ the original action \eqref{eq:S0-twoform} and $\Psi_\delta$ the gauge-fixing fermion \eqref{eq:psidelta}. This can be checked explicitly using formulas \eqref{eq:sbar}, or proven more abstractly as follows: since $\SNM$ is linear in antifields, we have $\SNM = S_0 + \Phi^*_I \frac{\delta^L \SNM}{\delta \Phi^*_I}$. Therefore,
\begin{align}
    S_\delta &= \SNM\left[ \Phi^I, \Phi^*_I = \frac{\delta \Psi_\delta}{\delta \Phi^I} \right] = S_0 + \frac{\delta \Psi_\delta}{\delta \Phi^I} \frac{\delta^L \SNM}{\delta \Phi^*_I} = S_0 + \bar{s}\, \Psi_\delta\, .
\end{align}
BRST invariance
\begin{equation}
    \bar{s}\, S_\delta = 0
\end{equation}
of the gauge-fixed action then follows from the gauge-invariance of $S_0$ (indeed, $\bar{s}S_0 = 0$ is equivalent to its gauge invariance since it only depends on $\psi_{\mu\nu}$) and $\bar{s}^2 = 0$.

We now do the same for the Gaussian gauge-fixing case. Even though \eqref{eq:sdelta} doesn't depend on the choice of gauge-fixing fermion, in the Gaussian case the non-minimal action from which we started is different. The BRST transformation in this case is then defined as
\begin{equation}\label{eq:stilde}
    \tilde{s} A = (A, \tSNM) = \frac{\delta^R A}{\delta \Phi^I}\frac{\delta^L \tSNM}{\delta \Phi^*_I}\, ,
\end{equation}
with $\tSNM$ given in \eqref{eq:SNMtilde-2form}. Again, this definition is valid because $\tSNM$ is linear in antifields, so no antifields appear on the right-hand-side of \eqref{eq:stilde}; otherwise, they should be eliminated using the gauge-fixing fermion $\Psi_\text{G}$. It takes the explicit form
\begin{align}
    \tilde{s}\,\psi_{\mu\nu} &= 2\, \pd_{[\mu} C_{\nu]}\, , \quad \tilde{s}\, C_\mu = \pd_\mu c\, , \quad \tilde{s}\,C'_\mu = b_\mu - \chi_\mu \nn \\
    \tilde{s}\,b_\mu &= 2\, \frac{\delta \chi_\mu}{\delta \psi_{\rho\sigma}} \pd_\rho C_\sigma \, , \quad \tilde{s}\, c' = \pi \, , \quad \tilde{s}\, \eta = \pi' \, , \quad \tilde{s}\, (\text{other}) = 0 \, .
\end{align}
Notice that $\tilde{s} \,b_\mu = \tilde{s} \,\chi_\mu$. The properties
\begin{align}
    \tilde{s}\,{}^2 = 0 \, , \quad
    S_\text{G} = S_0 + \tilde{s}\, \Psi_\text{G} \, , \quad
    \tilde{s}\, S_\text{G} = 0
\end{align}
then follow straightforwardly.

\paragraph{Propagators.}

We finish this section by exhibiting the propagators for $\psi_{\mu\nu}$ in both gauge-fixing schemes. The propagator $S^{\mu\nu}{}_{\sigma\tau}(p)$ is obtained by solving
\begin{equation}
    K^{\rho\kappa}{}_{\mu\nu}(p) S^{\mu\nu}{}_{\sigma\tau}(p) = \delta^{\rho\kappa}_{\sigma\tau}
\end{equation}
where $K^{\rho\kappa}{}_{\mu\nu}$ is the kinetic operator of $\psi_{\mu\nu}$ in momentum space.

In the delta-function gauge-fixing case \eqref{eq:Sgf-twoform-delta-final}, the kinetic part of the action for the gamma-traceless component $\hat{\psi}_{\mu\nu}$ of $\psi_{\mu\nu}$ is simply $-\slashed{\pd}$. The propagator for that component is therefore given by, including Feynman's $i\varepsilon$ prescription,
\begin{equation}
    S^{\mu\nu}{}_{\sigma\tau}(p) = - \mathbb{P}^{\mu\nu}{}_{\kappa\lambda}\, \frac{\slashed{p}}{p^2-i\varepsilon}\, \mathbb{P}^{\kappa\lambda}{}_{\sigma\tau}\, ,
\end{equation}
where $\mathbb{P}$ is the projector onto the gamma-traceless subspace
\begin{equation}
    \mathbb{P}^{\mu\nu}{}_{\rho\sigma} = \delta^{\mu\nu}_{\rho\sigma} + \frac{2}{D-2} \gamma^{[\mu} \delta^{\nu]}_{[\rho} \gamma_{\sigma]} - \frac{1}{(D-1)(D-2)} \gamma^{\mu\nu}\gamma_{\rho\sigma}\, .
\end{equation}
It is antisymmetric in both pairs of indices, satisfies $\gamma_\mu \mathbb{P}^{\mu\nu}{}_{\rho\sigma} = 0 = \mathbb{P}^{\mu\nu}{}_{\rho\sigma} \gamma^\rho$ and $\mathbb{P}^{\mu\nu}{}_{\kappa\lambda} \mathbb{P}^{\kappa\lambda}{}_{\rho\sigma} = \mathbb{P}^{\mu\nu}{}_{\rho\sigma}$.
The other component of $\psi_{\mu\nu}$ is $\rho$, which is simply a Dirac field with the usual propagator $-\slashed{p}/(p^2-i\varepsilon)$.

In the Gaussian gauge-fixing case, the kinetic operator appearing in \eqref{eq:Sgf-twoform-gaussian} reads, in momentum space,
\begin{align}
    K^{\rho\kappa}{}_{\mu\nu}(p) &= \gamma^{\rho\kappa\lambda}{}_{\mu\nu} p_\lambda - \xi \left( \delta^{[\rho}_\lambda \gamma^{\kappa]} + \frac{1}{D-2} \gamma_\lambda{}^{\rho\kappa} \right) \slashed{p} \left( \delta^{\lambda}_{[\mu} \gamma_{\nu]} - \frac{1}{D-2} \gamma^{\lambda}{}_{\mu\nu}\right)\, . 
\end{align}
We find the result
\begin{align}
    S^{\mu\nu}{}_{\sigma\tau}(p) =  \frac{1}{p^2 - i \varepsilon} \frac{1}{(D-4)} \Bigg[ & - \frac{1}{2} (D-4) \delta^{\mu\nu}_{\sigma\tau} \slashed{p} - 2 \left( p^{[\mu} \delta^{\nu]}_{[\sigma} \gamma_{\tau]} + \gamma^{[\mu} \delta^{\nu]}_{[\sigma} p_{\tau]} \right) + \gamma^{[\mu} \delta^{\nu]}_{[\sigma} \slashed{p} \gamma_{\tau]} \nn\\
    & + \frac{2}{D-2} \left( p^{[\mu}\gamma^{\nu]}\gamma_{\sigma\tau} + \gamma^{\mu\nu}\gamma_{[\sigma} p_{\tau]} \right) + \frac{1}{2(D-3)} \gamma^{\mu\nu}\slashed{p}\gamma_{\sigma\tau} \nn\\
    & + \frac{4}{(D-2)} \left( 1 + \frac{(D-4)[(D-2)^2+4] }{D^2\xi} \right) p^{[\mu} \gamma^{\nu]} \frac{\slashed{p}}{p^2} \gamma_{[\sigma}p_{\tau]} \nn\\
    & + 4 \left( 1 + \frac{(D-4)(D-2)^2}{D^2\xi} \right) p^{[\mu}  \delta^{\nu]}_{[\sigma} p_{\tau]} \frac{\slashed{p}}{p^2}\Bigg] \, .
\end{align}

\section{Gravitational anomalies}\label{sec:anomalies}

In this section, we compute the gravitational anomalies for chiral fermionic $p$-forms in $D = 4m + 2$ dimensions using the Atiyah-Singer index theorem \cite{Atiyah:1970ws,Atiyah:1971rm,Alvarez-Gaume:1983ihn,Alvarez-Gaume:1983ict,Alvarez-Gaume:1984zlq,Alvarez:1984yi}. More precisely, we will compute the anomaly polynomial as the $D+2$ form part of the index density of a Dirac operator,
\begin{equation}
    \hat{I}_{D+2} = [\text{Ind}(\slashed{D})]_{D+2}\, .
\end{equation}
The actual anomaly, which is a $D$-form, can be recovered from $\hat{I}_{D+2}$ by the method of descent (we refer to the review \cite{Bilal:2008qx} for further details).

As mentioned in the introduction, the computations of this section only rely on the spectrum of ghosts and they therefore apply to any theory with the same structure \eqref{eq:pformgauge} of gauge transformations and reducibilities, whether or not it has a kinetic term of the form \eqref{eq:S0-pform}. For this reason, we will also consider in this section fermionic $p$-forms with $2p \geq D$, which carry no degrees of freedom. For such fields, the action \eqref{eq:S0-pform} vanishes identically, but one could nevertheless imagine the existence of topological models in which they are coupled to other fields while still having the same structure of gauge transformations and reducibilities; our computations would then be applicable to such models. The prime example of this case is the gravitino in $D=2$, which doesn't have a Rarita-Schwinger kinetic term and carries no degree of freedom, but for which we nevertheless reproduce the classic result of \cite{Alvarez-Gaume:1983ihn}.

For definiteness, we take the field to be of positive chirality,
\begin{equation}
    \gamma_{*} \psi_{\mu_1\mu_2 \dots \mu_p} = + \psi_{\mu_1\mu_2 \dots \mu_p}\, ,
\end{equation}
where $\gamma_{*}$ is the usual chirality matrix. This then implies definite chiralities for all the other fields appearing in the gauge-fixed actions: these are found by simply requiring that no term in the gauge-fixed action vanish, remembering that a term of the form $\bar{\psi}_1 \gamma^{\mu_1 \dots \mu_n} \psi_2$ is non-zero when both $\psi_1$ and $\psi_2$ have the same chirality and $n$ is odd, or when $\psi_1$ and $\psi_2$ have opposite chiralities and $n$ is even. For example, the chiralities in the chiral two-form case are displayed in table \ref{tab:2formchiralities}.
\begin{table}
\centering
\begin{tabular}{c|cccccccccccc}
& $\psi_{\mu\nu}$ & $\hat{\psi}_{\mu\nu}$ & $\rho$ & $C_\mu$ & $C'_\mu$ & $c$ & $c'$ & $b_\mu$ & $\eta$ & $d_\mu$ & $\pi$ & $\pi'$  \\[\defaultaddspace] \midrule 
$S_\delta$ chirality & $+$ & $+$ & $+$ & $+$ & $+$ & $+$ & $+$ & $+$ & $+$ & $+$ & $+$ & $+$ \\[\defaultaddspace]
$S_\text{G}$ chirality & $+$ & & & $+$ & $-$ & $+$ & $+$ & $-$ & $-$ & & $+$ & $-$  \\[\defaultaddspace]
Grassmann parity & 1 & 1 & 1 & 0 & 0 & 1 & 1 & 1 & 1 & 1 & 0 & 0
\end{tabular}
\caption{The chirality of the various fields appearing in the gauge-fixed actions \eqref{eq:Sgf-twoform-delta-final} and \eqref{eq:Sgf-twoform-gaussian} for the chiral fermionic two-form. Notice that it can depend on the gauge-fixing scheme. Grassmann parity is also included; even fields have abnormal spin-statistics.}
\label{tab:2formchiralities}
\end{table}

We work in this section with a (spin, orientable) manifold $\cM$ of Euclidean signature (for a detailed account of the continuation from Minkowskian to Euclidean signature, see e.g.~\cite{Bilal:2008qx}). The space of fermionic $p$-forms on $\cM$ will be written as $\mathcal{C}^{\infty}(S \otimes \Lambda^p \,T^*\!\cM)$, where $S$ is the relevant spinor representation of $SO(D)$.

The anomaly polynomial is given in terms of the index density of a certain Dirac operator \cite{Alvarez-Gaume:1983ict,Alvarez-Gaume:1984zlq}. Following \cite{Alvarez:1984yi}, we will write the total Dirac operator, including all fields, in the form
\begin{equation} \label{eq:Diracstandard}
\slashed{D} : \mathcal{C}^{\infty} (S^+ \otimes \cV) \longrightarrow \mathcal{C}^{\infty}(S^- \otimes \cV)\, ,
\end{equation}
where $S^+$ (resp. $S^-$) denotes the positive (resp. negative) spinor representation of $SO(D)$, and $\cV$ is a formal sum of spaces carrying tensor representations of $SO(D)$. This means that the corresponding index densities should be added or subtracted according to the signs appearing in the formal sum $\cV$.

Some formal manipulations and careful handling of signs are required to reach the standard form \eqref{eq:Diracstandard} and compute the relevant $\cV$ \cite{Alvarez:1984yi}. Let us show how they work explicitly in the chiral two-form case. The computation is slightly different in the two gauge-fixing schemes of this paper, but the result is of course the same.
\paragraph{Delta-function gauge-fixing.} After integrating out the auxiliary fields, the relevant path integral measure reads (omitting the complex conjugates)
\begin{equation}\label{eq:measuredelta}
    \int \cD \hat{\psi}_{\mu\nu} \,\cD \rho \,\cD \hat{C}_\mu \, \cD \hat{C}'_\mu \,\cD c \,\cD c'
\end{equation}
where a `hat' denotes a gamma-traceless field. Since the gamma-trace has the opposite chirality from the field itself, the field $\hat{\psi}_{\mu\nu}$ for example can be seen as an element of the formal difference
\begin{equation}\label{eq:psihatcomplex}
    \mathcal{C}^{\infty}(S^+ \otimes \Lambda^2 T^*\!\cM - S^- \otimes T^*\!\cM)\, ,
\end{equation}
i.e.~a positive chirality fermionic two-form without the negative chirality one-form component. This is not in the standard form \eqref{eq:Diracstandard} for $\slashed{D}$ to act upon; however, a fermion of negative chirality gives the opposite contribution to the index density as a fermion of positive chirality. We can then replace $S^-$ by $S^+$ in \eqref{eq:psihatcomplex} and change the sign: $\hat{\psi}_{\mu\nu}$ therefore contributes as \begin{equation}
\mathcal{C}^{\infty}(S^+ \otimes [\Lambda^2 T^*\!\cM + T^*\!\cM])\, ,    
\end{equation}
which is now in the form \eqref{eq:Diracstandard}. Another rule is that fields with the wrong spin-statistics, in our case $\hat{C}_\mu$ and $\hat{C}'_\mu$, also contribute with a minus sign. Combining these two rules, the (wrong spin-statistics, positive chirality, gamma-traceless) field $\hat{C}_\mu$ for example contributes as
\begin{align}
    - \mathcal{C}^{\infty}(S^+ \otimes T^*\!\cM - S^-) &= \mathcal{C}^{\infty}(S^+ \otimes [ - T^*\!\cM - 1])\, .
\end{align}
One must now sum the contributions of all fields appearing in the measure \eqref{eq:measuredelta}, using these two rules and the chirality and spin-statistics of table \ref{tab:2formchiralities}. The complex $\mathcal{C}^{\infty} (S^+ \otimes \cV_\delta)$ on which the Dirac operator acts in this case is then
\begin{align}
\mathcal{C}^{\infty} (S^+ \otimes \cV_\delta) &= \mathcal{C}^{\infty}(S^+ \otimes \Lambda^2 T^*\!\cM - S^- \otimes T^*\!\cM) + \mathcal{C}^{\infty}(S^+) \nn\\
&\quad - 2\, \mathcal{C}^{\infty}(S^+ \otimes T^*\!\cM - S^-) + 2\, \mathcal{C}^{\infty}(S^+) \\
    &= \mathcal{C}^{\infty}(S^+ \otimes [\Lambda^2 T^*\!\cM - T^*\!\cM + 1])\nn
\end{align}
and we have
\begin{equation}
    \cV_\delta = \Lambda^2 T^*\!\cM - T^*\!\cM + 1\, .
\end{equation}

\paragraph{Gaussian gauge-fixing.}
In the Gaussian gauge-fixing case, the Nielsen-Kallosh ghost $b_\mu$ enters the dynamics and we have the measure 
\begin{equation}
    \int \cD \psi_{\mu\nu} \, \cD \hat{b}_\mu \,\cD \hat{C}_\mu \, \cD \hat{C}'_\mu \,\cD c \,\cD c'
\end{equation}
after integrating out the auxiliary fields. Here as before, a hat indicates a gamma-traceless field. Notice that in this case (cf.~table \ref{tab:2formchiralities}), the ghosts $\hat{C}_\mu$ and $\hat{C}'_\mu$ have opposite chiralities but otherwise identical properties and their contributions to the index density cancel out. The total complex on which $\slashed{D}$ acts is then
\begin{align}
\mathcal{C}^{\infty} (S^+ \otimes \cV_\text{G}) &= \mathcal{C}^{\infty}(S^+ \otimes \Lambda^2 T^*\!\cM ) + \mathcal{C}^{\infty}(S^- \otimes T^*\!\cM - S^+) + 2\, \mathcal{C}^{\infty}(S^+)  \nn\\
    &= \mathcal{C}^{\infty}(S^+ \otimes [\Lambda^2 T^*\!\cM - T^*\!\cM + 1])\, .
\end{align}
Therefore,
\begin{equation}
    \cV_{\text{G}} = \cV_{\delta} \equiv \cV_2 
\end{equation}
as expected; both gauge-fixing procedures will give the same result for the anomaly.

It is now an easy exercise to repeat this procedure for the higher $p$-form cases, using the information of sections \ref{sec:BVredhigherstage} and \ref{sec:pformactionetc}. For the three-form we find $\cV_3 = \Lambda^3 T^*\!\cM - \Lambda^2 T^*\!\cM + T^*\!\cM - 1$; more generally, one has
\begin{equation}\label{eq:complex}
    \cV_p = \sum_{k=0}^p (-1)^k\, \Lambda^{p-k} \, T^*\!\cM\, .
\end{equation}

With $\slashed{D}$ in the standard form, its index density is then \cite{Atiyah:1970ws,Atiyah:1971rm,Alvarez:1984yi}
\begin{equation}
\label{eq:IDirac}
\text{Ind}(\slashed{D}) = \AM \, \ch(R,\cV) \, .
\end{equation}
The first factor $\AM$ in this formula is known as the Dirac genus (or roof genus) of $\cM$. It is given in terms of the curvature $2$-form $R^a{}_b$ of $\cM$ by
\begin{align}
\AM = 1 &+ \frac{1}{(4 \pi)^2} \frac{1}{12} \tr(R^2)+\frac{1}{(4 \pi)^4} \left[\frac{1}{288} \tr(R^2)^2 + \frac{1}{360} \tr(R^4)\right] \nn \\
&+\frac{1}{(4 \pi)^6} \left[ \frac{1}{10368} \tr(R^2)^3 + \frac{1}{4320} \tr(R^2)\tr(R^4) + \frac{1}{5670} \tr(R^6) \right]+\dots
\end{align}
Here, the traces are taken in the fundamental (vector) representation of $SO(D)$. Writing components explicitly, with $R^a{}_b = \frac{1}{2} R^a{}_{b\, \mu\nu} dx^\mu dx^\nu$, one has for example $\tr(R^2) = R^a{}_b R^b{}_a = \frac{1}{4} R^a{}_{b\, \mu\nu} R^b{}_{a\, \rho\sigma} \,dx^\mu dx^\nu dx^\rho dx^\sigma$. The genus $\AM$ is thus a sum of differential forms of degrees that are multiples of four. It will be useful to write the results in terms of Pontryagin classes $p_i$ rather than the curvature itself; they are defined by the expansion
\begin{equation}
\det\left(1-\frac{R}{2\pi}\right) = 1 + p_1 + p_2 + p_3 +  p_4+\dots
\end{equation}
with each $p_i$ a form of degree $4i$. The first three Pontryagin classes will be sufficient for our purposes:
\begin{align}
p_1 &= \frac{1}{(2 \pi)^2} \left(-\frac{1}{2} \tr R^2 \right)\\
p_2 &=\frac{1}{(2 \pi)^4} \left(-\frac{1}{4} \tr R^4 +\frac{1}{8} (\tr R^2)^2 \right) \\
p_3 &= \frac{1}{(2 \pi)^6} \left(-\frac{1}{6} \tr R^6 +\frac{1}{8} \tr R^2 \tr R^4 - \frac{1}{48} (\tr R^2)^3 \right) \, .
\end{align}
The Dirac genus can then be written as
\begin{align}
\AM &= 1- \frac{1}{24} \,p_1 + \frac{1}{5760}(7 \, p^2_1 - 4 \, p_2) + \frac{1}{967680} \left( -31 \, p_1^3 + 44 \, p_1 p_2 - 16 \, p_3 \right) + \dots\, .
\end{align}

The second factor in \eqref{eq:IDirac} is the Chern character, which for a single representation $\mathfrak{r}$ of $SO(D)$ reads
\begin{equation}
\text{ch}(R,\mathfrak{r}) = \Tr \exp\left( \frac{i R_\mathfrak{r}}{2\pi} \right)\, .
\end{equation}
In this formula, the two-form $R_\mathfrak{r}$ is defined as $R_\mathfrak{r} = \frac{1}{2}R_{ab} T_\mathfrak{r}^{ab}$, where $R_{ab}$ is the curvature two-form of $\cM$ and the $T_\mathfrak{r}^{ab}$ are the $SO(D)$ generators for the representation $\mathfrak{r}$. The traces are taken in the representation at hand. When $\cV$ is a formal sum of representation spaces, as in \eqref{eq:complex}, one takes the corresponding sum of characters. The representations that appear in this case are the antisymmetric tensor representations of $SO(D)$, which we write as $\mathfrak{r} = [k]$ for some integer $k$. Their Chern character is
\begin{equation}
\ch(R,[k]) = \Tr \exp\left( \frac{i R_{[k]}}{2\pi} \right) = \frac{D!}{k!(D-k)!} - \frac{1}{2!} \frac{1}{(2\pi)^2} \Tr (R_{[k]}^2) + \frac{1}{4!} \frac{1}{(2\pi)^4} \Tr(R_{[k]}^4) + \dots \, ,
\end{equation}
where we used $\Tr(\id) = \dim([k]) = \frac{D!}{k!(D-k)!}$ and the fact that traces of odd powers of $R_{[k]}$ identically vanish for these representations. The generators $T_{[k]}^{ab}$ are given by
\begin{align}\label{eq:Tk}
(T_{[k]}^{ab})^{i_1 \dots i_k}{}_{j_1 \dots j_k} = k! \sum_{l=1}^k \delta^{[i_1}_{j_1} \delta^{i_2}_{j_2} \cdots (t^{ab})^{i_l}{}_{j_l} \cdots \delta^{i_k]}_{j_k}\, ,
\end{align}
with $(t^{ab})^i{}_j = 2 \delta^{i[a} \delta^{b]}_j$ the generators of the fundamental. Therefore, $R_{[k]}$ is
\begin{align}\label{eq:Rk}
(R_{[k]})^{i_1 \dots i_k}{}_{j_1 \dots j_k} &= \frac{1}{2} R_{ab} (T_{[k]}^{ab})^{i_1 \dots i_k}{}_{j_1 \dots j_k} = k! \sum_{l=1}^k \delta^{[i_1}_{j_1} \delta^{i_2}_{j_2} \cdots R^{i_l}{}_{j_l} \cdots \delta^{i_k]}_{j_k}\, .
\end{align}
The next step of the computation is to write the traces of powers of $R_{[k]}$ in terms of Pontryagin classes or, equivalently, in terms of traces of powers of $R$ in the fundamental representation. For low $k$ and low powers this can be done using the explicit formula \eqref{eq:Rk}: for example, for $k=2$ one has $R_{[2]}{}^{ij}{}_{kl} = 2 R^{[i}{}_k \delta^{j]}_l + 2 \delta^{[i}_k R^{j]}{}_l$ and then
\begin{align}
    \Tr (R_{[2]}^2) &= \frac{1}{4} R_{[2]}{}^{ij}{}_{kl} R_{[2]}{}^{kl}{}_{ij} = (D-2) R^i{}_j R^j{}_i = (D-2)\tr(R^2)\, , \label{eq:TrR22}\\
    \Tr (R_{[2]}^4) &= \frac{1}{16} R_{[2]}{}^{ij}{}_{kl} R_{[2]}{}^{kl}{}_{mn} R_{[2]}{}^{mn}{}_{pq} R_{[2]}{}^{pq}{}_{ij} = (D-8) \tr(R^4) + 3 (\tr R^2)^2\, . \label{eq:TrR24}
\end{align}
This quickly becomes computationally intractable, however. Fortunately, this is a well-known mathematical problem and there exists the following generating formula (see e.g.~\cite{Schellekens:1986xh}):
\begin{align}\label{eq:generatingformula}
\sum_{k=0}^\infty x^k \ch(R,[k]) = \det\left(1 + x \, e^{\frac{i R}{2\pi}}\right) = \exp \tr \log \left(1 + x \, e^{\frac{i R}{2\pi}}\right)
\end{align}
with $x$ a formal variable. A formula for $\ch(R,[k])$ can be extracted by expanding the right-hand-side as a formal power series in $x$ and selecting the coefficient of $x^k$. For example, from the coefficient of $x^2$ one finds
\begin{equation}
\ch(R,[2]) = \frac{1}{2} \left( \tr e^{\frac{i R}{2\pi}} \right)^2 - \frac{1}{2} \tr e^{\frac{i 2 R}{2\pi}}\,.
\end{equation}
Expanding the exponentials, this formula contains all $\Tr (R_{[2]}^n)$ in terms of traces in the fundamental: the four-form component of this equation reproduces equation \eqref{eq:TrR22}, the eight-form component gives \eqref{eq:TrR24}, and so on. Likewise, any trace $\Tr (R_{[k]}^n)$ can be found by expanding equation \eqref{eq:generatingformula} to order $x^k$ and to form degree $2n$.

Finally, the anomaly polynomial for a chiral fermionic $p$-form is given as the $D+2$ form part in the index density \eqref{eq:IDirac}. Using the form \eqref{eq:complex} of $\cV_p$, this is
\begin{equation}
\hat{I}^{(p)}_{D+2} = \left[ \AM \sum_{k=0}^p (-1)^{p-k} \ch(R,[k]) \right]_{D+2}\, .
\end{equation}
This can be computed for any desired $D$ and $p$ using the ingredients detailed above. We display explicitly the results in terms of Pontryagin classes in dimensions $D = 2$, $6$ and $10$ in tables \ref{tab:D2}, \ref{tab:D6} and \ref{tab:D10}. Of course, for spin $1/2$ and $3/2$ fields ($p=0$ and $p=1$ respectively), these tables reproduce the classic results of \cite{Alvarez-Gaume:1983ihn}. The anomaly polynomial for the chiral bosons (i.e.~the self-dual scalar, $2$-form and $4$-form) in those dimensions are also listed for convenience \cite{Alvarez-Gaume:1983ihn}.

Interestingly, in dimensions $D \geq 6$ we find\footnote{To be more precise: this is apparent in $D=6$ and $10$ from tables \ref{tab:D6} and \ref{tab:D10}, and has been checked explicitly in $D=14$ and $18$; however, we have no general proof for arbitrary $D$.} that the anomaly of a chiral fermionic $p$-form matches that of a $(D-p-1)$-form,
\begin{equation}\label{eq:anomalyrelation}
\hat{I}^{(p)}_{D+2} = \hat{I}^{(D-p-1)}_{D+2}\, .
\end{equation}
For example, the anomaly of a chiral fermionic $2$-form in $D=6$ could be cancelled by a $3$-form of the opposite chirality. Similarly, one could imagine canceling the anomaly of a bosonic, self dual $4$-form in $D=10$ such as the one appearing in type IIB supergravity using topological fermionic $8$- and $9$-forms of opposite chirality. (Notice how \eqref{eq:anomalyrelation} always relates the anomaly polynomial of a dynamical field to that of a topological one.) This is of course subject to the caveats mentioned in the introduction, namely, the current lack of explicit Lagrangians coupling fermionic $p$-forms to dynamical gravity. Nevertheless, it would be very interesting to see whether these possibilities can be realised in physically relevant models.

\begin{table}
    \centering
    \begin{tabular}{c| l}
        $p$ & $\hat{I}^{(p)}_{4}$ \\ \midrule
        0 & $-\frac{1}{24} \,p_1$ \\[\defaultaddspace]
        1 & $\frac{23}{24} \,p_1$ \\[\defaultaddspace]
        2 & $- p_1$ \\[\defaultaddspace] \midrule
        $\hat{I}^A_{4}$ & $-\frac{1}{24} \,p_1$
    \end{tabular}
    \caption{The four-form anomaly polynomials for chiral fermionic $p$-forms in $D=2$.}
    \label{tab:D2}
\end{table}

\begin{table}
    \centering
    \begin{tabular}{c| l}
        $p$ & $\hat{I}^{(p)}_{8}$ \\ \midrule
        0 & $\frac{1}{5\,760} \left( 7 \,p^2_1 - 4 \,p_2 \right)$ \\[\defaultaddspace]
        1 & $\frac{1}{5\,760} \left( 275 \,p^2_1 - 980 \,p_2 \right)$ \\[\defaultaddspace]
        2 & $\frac{1}{5\,760} \left( 790 \,p^2_1 + 2\,840 \,p_2 \right)$ \\[\defaultaddspace]
        3 & $\frac{1}{5\,760} \left( 790 \,p^2_1 + 2\,840 \,p_2 \right)$ \\[\defaultaddspace]
        4 & $\frac{1}{5\,760} \left( 275 \,p^2_1 - 980 \,p_2 \right)$ \\[\defaultaddspace]
        5 & $\frac{1}{5\,760} \left( 7 \,p^2_1 - 4 \,p_2 \right)$ \\[\defaultaddspace]
        6 & $0$ \\[\defaultaddspace] \midrule
        $\hat{I}^A_{8}$ & $\frac{1}{5\,760} \left( 16 \,p^2_1 - 112 \,p_2 \right)$
    \end{tabular}
    \caption{The eight-form anomaly polynomials for chiral fermionic $p$-forms in $D=6$.}
    \label{tab:D6}
\end{table}

\begin{table}
    \centering
    \begin{tabular}{c| l}
        $p$ & $\hat{I}^{(p)}_{12}$ \\ \midrule
        0 & $\frac{1}{967\,680} \left( -31 \,p_1^3 + 44 \,p_1 p_2 - 16 \,p_3 \right)$ \\[\defaultaddspace]
        1 & $\frac{1}{967\,680} \left( 225 \,p_1^3 - 1\,620 \,p_1 p_2 + 7\,920 \,p_3 \right)$ \\[\defaultaddspace]
        2 & $\frac{1}{967\,680} \left( 2\,412 \,p_1^3 + 27\,792 \,p_1 p_2 - 186\,048 \,p_3 \right)$ \\[\defaultaddspace]
        3 & $\frac{1}{967\,680} \left( 7\,980 \,p_1^3 + 162\,960 \,p_1 p_2 - 73\,920\, p_3 \right)$ \\[\defaultaddspace]
        4 & $\frac{1}{967\,680} \left( 13\,734 \,p_1^3 + 338\,184\, p_1 p_2 + 764\,064 \,p_3 \right)$ \\[\defaultaddspace]
        5 & $\frac{1}{967\,680} \left( 13\,734 \,p_1^3 + 338\,184 \,p_1 p_2 + 764\,064 \,p_3 \right)$ \\[\defaultaddspace]
        6 & $\frac{1}{967\,680} \left( 7\,980\, p_1^3 + 162\,960 \,p_1 p_2 - 73\,920 \,p_3 \right)$ \\[\defaultaddspace]
        7 & $\frac{1}{967\,680} \left( 2\,412 \,p_1^3 + 27\,792\, p_1 p_2 - 186\,048 \,p_3 \right)$ \\[\defaultaddspace]
        8 & $\frac{1}{967\,680} \left( 225\, p_1^3 - 1\,620\, p_1 p_2 + 7\,920\, p_3 \right)$ \\[\defaultaddspace]
        9 & $\frac{1}{967\,680} \left( -31\, p_1^3 + 44\, p_1 p_2 - 16\, p_3 \right)$ \\[\defaultaddspace]
        10 & $0$ \\[\defaultaddspace] \midrule
        $\hat{I}^A_{12}$ & $\frac{1}{967\,680} \left( -256\, p_1^3 + 1\,664\, p_1 p_2 - 7\,936 \,p_3 \right)$
    \end{tabular}
    \caption{The twelve-form anomaly polynomials for chiral fermionic $p$-forms in $D=10$.}
    \label{tab:D10}
\end{table}

\subsubsection*{Acknowledgements}

We would like to thank Alex S. Arvanitakis, Marc Henneaux, Chris Hull, Axel Kleinschmidt and Ruben Minasian for useful discussions. Yi Zhang would like to thank the Max-Planck-Institut für Gravitationsphysik (Albert-Einstein-Institut) for hospitality. 

The work of VL is supported by the European Research Council (ERC) under the European Union’s Horizon 2020 research and innovation programme, grant agreement No. 740209. The work of YZ is supported by the International Max Planck Research School for Mathematical and Physical Aspects of Gravitation, Cosmology and Quantum Field Theory.


\providecommand{\href}[2]{#2}\begingroup\raggedright\endgroup

\end{document}